\documentclass[twocolumn,amsmath,amssymb,floats,superscriptaddress,nofootinbib,10pt]{revtex4-1}

\newcommand{\quotes}[1]{``#1''}

\newcommand{\Rhat}{\widehat R}
\newcommand{\OmF}{\Omega_{\rm F}}
\newcommand{\dd}{\mathrm d}
\newcommand{\Nlapse}{\alpha  }
\newcommand{\DtPhi}{\mathcal D_t^{(\Phi)}}

\pdfoutput=1

\AtBeginDocument{%
    \newwrite\bibnotes
    \def\bibnotesext{Notes.bib}
    \immediate\openout\bibnotes=\jobname\bibnotesext
    \immediate\write\bibnotes{@CONTROL{%
    apsrev41Control,author="08",editor="1",pages="1",title="0",year="1"}}
     \if@filesw
     \immediate\write\@auxout{\string\citation{apsrev41Control}}%
    \fi
}%

\usepackage{graphicx}
\usepackage{dcolumn}
\usepackage{bm}
\usepackage{revsymb}
\usepackage[usenames]{color}
\usepackage{color}
\usepackage{physics}
\usepackage{amsfonts}
\usepackage[section]{placeins} 

\usepackage{xcolor}
\usepackage{url}
\usepackage{tikz}

\usepackage{dblfloatfix} 

\usepackage{amsmath}
\usepackage{ulem}
\usepackage[usenames]{color}
\usepackage{amsfonts}
\usepackage{graphicx}
\usepackage{dcolumn}
\usepackage{bm}
\usepackage{revsymb}
\usepackage[usenames]{color}
\usepackage{graphicx}

\usepackage{color}
\usepackage{physics}
\usepackage{amsmath}
\usepackage{amssymb,bbm}
\usepackage{lipsum}  
\usepackage{float}

\usepackage{subcaption}

\usepackage[version=4]{mhchem}

\usepackage{xcolor}

\begin{document}

\title{Resistive alignment of a black-hole magnetosphere}

\author{Ruben Lier}
\email{r.lier@uva.nl}
\affiliation{Institute for Theoretical Physics, University of Amsterdam, 1090 GL Amsterdam, The Netherlands}
\affiliation{Dutch Institute for Emergent Phenomena (DIEP), University of Amsterdam, 1090 GL Amsterdam, The Netherlands}
\affiliation{Institute for Advanced Study, University of Amsterdam, Oude Turfmarkt 147, 1012 GC Amsterdam, The Netherlands}
\author{Oliver Porth}\email{o.j.g.porth@uva.nl}
\affiliation{Anton Pannekoek Institute, Science Park 904, 1098 XH, Amsterdam, The Netherlands}

\begin{abstract}
Recent numerical simulations have shown that black-hole magnetospheres evolve rapidly toward alignment between the magnetic axis and spin axis but the physical origin of this process has remained unclear. We develop an analytic theory of this alignment within resistive relativistic magnetohydrodynamics. Taking the slow-spin Blandford–Znajek monopole solution as the exterior magnetic field, we reduce the induction equation to a relaxation equation for the current-sheet inclination angle.  Our analysis shows that alignment results from second-order spin effects in the Blandford–Znajek solution combined with explicit resistive dissipation. We test this prediction using resistive-GRMHD simulations at several spins and resistivities and compare the measured late-time secular decay with the analytic prediction.
This alignment timescale hence carries a unique signature which combines plasma dissipation with spacetime rotation.
\end{abstract}

\maketitle

\section{Introduction}
Black-hole--neutron-star mergers can leave behind a transient black-hole magnetosphere, even when the neutron star is swallowed without tidal disruption \cite{Kim2025}. Pair creation supplies the charges and currents needed to sustain this magnetosphere outside the horizon \cite{Levinson2018,PhysRevLett.124.145101}.  If the black hole
rotates, this inherited magnetic flux can extract rotational energy
through the Blandford--Znajek mechanism \cite{Blandford1977}.  The resulting magnetosphere is initially out of equilibrium and must subsequently relax toward a stationary split-monopolar structure \cite{PhysRevD.84.084019,PhysRevLett.127.055101,Selvi2025}. In particular, the magnetosphere can initially be tilted with respect to the spin axis as recently demonstrated in simulations of black-hole--neutron-star mergers \cite{Kim2025}.  Simulations have provided empirical evidence for the subsequent secular alignment of the magnetic and spin axes on a rapid timescale $O(100) GM/c^3$ that depends on the magnitude of the black hole spin as $a^{-2}$ \cite{Selvi_2024} but its physical origin and resistive scaling have remained unclear.

Two examples of spin alignment in other astrophysical contexts are the Bardeen-Petterson alignment of accretion disks \citep{BardeenPetterson1975,PapaloizouPringle1983} and the secular alignment of pulsar magnetospheres \cite{MichelGoldwire1970,Philippov_2014}.  
In the Bardeen-Petterson (BP) effect, the angular momentum of a thin accretion disk is brought into alignment with the black hole spin, caused by differential spacetime rotation and viscous transport of mis-aligned angular momentum.  
For pulsars with dipolar magnetic fields, alignment is caused by magnetospheric torques and occurs on the spindown time of $\sim$ Gyr.  \cite{Philippov_2014}.
In contrast, the alignment torque vanishes for the ideal force-free split-monopole \cite{Bogovalov:1999vg} -- which resembles the black hole magnetospheric geometry. However, a black hole horizon substantially differs from the neutron star surface which can be approximated as a perfectly conducting sphere.  Thus instead of being \quotes{frozen in} to the rotating conductor, field lines crossing the event horizon are free to slide across, subject to frame-dragging and electromagnetic forces \cite{Selvi2025}.  

In ideal force-free electrodynamics, the split-monopole magnetosphere is locally identical to the monopolar magnetosphere away from the current sheet: the two hemispheres differ only by a reversal of electromagnetic polarity. The distinction between a monopole and a split monopole therefore resides entirely in the magnetic-reversal surface, and ideal force-free electrodynamics by itself provides no mechanism that selects its aligned configuration. A split-monopole magnetosphere, however, necessarily contains a current sheet at this reversal surface, where the ideal description breaks down and magnetic reconnection enters. The natural question is therefore whether non-ideal effects drive the current sheet toward alignment and, if so, what sets the corresponding rate.

Here we show that magnetic diffusion alone produces a systematic
alignment of a weakly inclined Blandford-Znajek current sheet.  By
tracking the magnetic-reversal surface, we derive the
horizon alignment rate
\begin{equation}  \label{eq:mainresult}
 \Gamma_\chi
 =   0.58544\,
 \frac{\eta_{\rm m}a_\star^2}{M^2}
 +O(\eta_{\rm m}a_\star^4/M^2),
\end{equation}
predicting a rate linear in magnetic resistivity $\eta_{\rm m }$ and quadratic in dimensionless
black-hole spin $a_{\star}$. We construct the corresponding inclined Blandford--Znajek split monopole, including the radial warp induced by the \(O(a_\star^2)\) deformation of the magnetic flux surfaces, and test the predicted alignment law with three-dimensional resistive-GRMHD simulations. The simulations allow us to compare the late-time alignment with the predicted resistivity and spin dependencies of Eq.~\eqref{eq:mainresult}, and to examine departures at low and high resistivity.

\section{Resistive Magnetohydrodynamics}

We describe the magnetosphere using resistive general-relativistic
magnetohydrodynamics in geometric units $G=c=1$ and
Heaviside--Lorentz units. The dynamical equations are
\begin{align}  \label{eq:main_mass}
\begin{split}
     \nabla_\mu(\rho u^\mu)&=0,
 \qquad  \nabla_\mu T^{\mu\nu}=0,
\\
 \nabla_\nu {}^\star F^{\mu\nu}&=0,
 \qquad
 \nabla_\nu F^{\mu\nu}=\mathcal J^\mu ,
 \end{split}
\end{align}
where $\rho$ is the fluid rest-mass density and $u^\mu$ its
four-velocity. The total stress tensor is
\begin{equation}
 T^{\mu\nu}
 =
 \rho h u^\mu u^\nu
 +p g^{\mu\nu}
 +F^{\mu\lambda}F^\nu{}_{\lambda}
 -\frac14 g^{\mu\nu}F^{\rho\sigma}F_{\rho\sigma},
 \label{eq:main_Tmunu}
\end{equation}
with pressure $p$ and specific enthalpy $h$. To close the electromagnetic sector, we decompose the four-current
with respect to the Eulerian observer $n^\mu$ as
\begin{equation}
 \mathcal J^\mu=q n^\mu+J^\mu,
 \qquad
 q=-n_\mu\mathcal J^\mu,
 \qquad
 n_\mu J^\mu=0 ,
 \label{eq:main_current_split}
\end{equation}
where $q$ is the Eulerian charge density and $J^i$ the spatial
electric current. The electric and magnetic fields measured by the Eulerian observer are
\[
E^\mu = F^{\mu\nu}n_\nu,
~~ 
B^\mu = {}^\star F^{\mu\nu}n_\nu,
~~ n_\mu E^\mu=n_\mu B^\mu=0,
\]
with spatial components \(E^i\) and \(B^i\). Writing
$u^\mu=\Gamma(n^\mu+v^\mu)$, we adopt the scalar resistive Ohm's law \cite{PhysRevD.88.044020}
\begin{equation}
 J^i
 =
 qv^i
 +\frac{\Gamma}{\eta_{\rm m}}
 \left[
 E^i+\epsilon^{ijk}v_jB_k
 -(E^jv_j)v^i
 \right]   , 
 \label{eq:main_Ohm}
\end{equation}
where \(\epsilon^{ijk}\) is the spatial Levi–Civita tensor, with \(\epsilon^{123}=1/\sqrt{\gamma}\). Equations~\eqref{eq:main_mass}--\eqref{eq:main_Ohm} constitute the
full system used in the numerical calculations below. For the analytic calculation, we subsequently take the
quasistatic magnetic-diffusion limit of this system.

To arrive at the induction equation that allows for finding the alignment time, as detailed in the Supplementary Material, we neglect the
displacement-current contribution to Amp\`ere's law \cite{Geppert_2000,Rezzolla_2001}, charge
advection in the non-ideal current, the longitudinal electric-field
term, and relativistic corrections to the resistive response.  This
gives the curved-space induction equation 
\begin{equation}
 \begin{aligned}
  \partial_t(\sqrt\gamma B^i)
  & +\partial_j\Big\{\sqrt\gamma\Big[
 V^jB^i-V^iB^j\\
 &+\eta_{\rm m}
 \big(D^i(\Nlapse B^j)-D^j(\Nlapse B^i)\big)
 \Big]\Big\}=0 .
 \end{aligned}
 \label{eq:main_induction}
\end{equation}
with the transport velocity
\begin{equation}
 V^i=\Nlapse v^i-\beta^i 
\end{equation}
in 3+1 decomposition \cite{Arnowitt2008}.

\section{Evolution of inclination angle $\chi$}
We now use Eq.~\eqref{eq:main_induction} to obtain an evolution equation of the inclination angle $\chi$ of the current sheet. To do this, we must assume that the sheet is a common
reversal of the magnetic field,
\begin{equation}
 B^i
 =
 \mathcal Q(\Xi)B_{\rm ext}^i ,
 \qquad
 \mathcal Q(0)=0 ,
 \qquad
 \mathcal Q'(0)\neq0 ,
 \label{eq:main_common_reversal}
\end{equation}
where $\mathcal Q\in[-1,1]$ is a smooth reversal profile, $\Xi=\Xi_{\rm BZ}(\psi,\Phi;\chi)$ is the field-line label defined in Eq.~\eqref{eq:main_exact_sheet}, $\Xi=0$ defines the sheet $\Sigma$, and $B_{\rm ext}^i$ is the smooth exterior field. Any constant normalization of the transition width is absorbed into $\mathcal Q$; primes denote derivatives with respect to its argument. Because $\Xi$ is constant along the exterior magnetic field on each spatial slice, this smooth reversal satisfies the magnetic constraint throughout the layer to the retained order.  Thus all magnetic-field components vanish on
the same surface.  Since $B_{\rm ext}^r$ is nonzero in the monopolar
exterior, the complete reversal surface may equivalently be tracked
by 
\begin{equation}
 {\cal B}\equiv B^r,
 \qquad
 {\cal B}|_\Sigma=0 .
 \label{eq:main_marker}
\end{equation}
Differentiating
${\cal B}(t,r,\theta_\Sigma,\phi)=0$ gives \cite{murphymurphy,parnell,10.1063/1.860718}
\begin{equation}
 \left[
 \DtPhi{\cal B}
 +(\DtPhi\theta_\Sigma)\partial_\theta{\cal B}
 \right]_\Sigma  =0 ,
~~ \DtPhi\equiv\partial_t+\OmF\partial_\phi .
 \label{eq:main_levelset}
\end{equation}
Substitution of the radial component of
Eq.~\eqref{eq:main_induction} therefore directly determines the
velocity of the magnetic zero. As shown in the Supplementary Material, taking its $m=1$ azimuthal Fourier amplitude gives
\begin{equation}
 \frac{\dd\chi}{\dd t}
 =
 {\mathfrak A}_{\rm pl}
 +{\mathfrak R}_{\rm diff},
 \label{eq:main_chi_ledger}
\end{equation}
where ${\mathfrak A}_{\rm pl}$ contains advection by the unresolved
plasma flow and ${\mathfrak R}_{\rm diff}$ is determined by magnetic
diffusion.

The term ${\mathfrak R}_{\rm diff}$ provides a resistive alignment rate. To determine it requires knowledge of an exterior field, which we take to be the stationary
slow-spin Blandford--Znajek (BZ) monopole \cite{Blandford1977,McKinney_2004}. In horizon-regular ingoing Kerr--Schild coordinates it can be written, to the required order, as \cite{tanabe,Armas_2020,GrallaJacobson2014}
\begin{equation}
 F=\dd\psi\wedge\dd\Phi+O(a_\star^3),
 \label{eq:main_BZ_euler}
\end{equation}
with
\begin{equation}
 \psi
 =
 \psi_0(1-\cos\theta)
 +a_\star^2\psi_0\Rhat(r/M)
     \sin^2\theta\cos\theta
 +O(a_\star^4),
 \label{eq:main_BZ_flux}
\end{equation}
and
\begin{align}
 \OmF  & =\frac{a_\star}{8M}+O(a_\star^3/M),
  \nonumber  \\ 
 \Phi
   & =
 \phi-\OmF\left[
 t-r-4M\ln\!\left(\frac{r}{2M}\right)
 \right]-\Phi_0 .
 \label{eq:main_BZ_phase}
\end{align}
Here $a_\star=a/M$, while the explicit radial function $\Rhat$ is
given in the Supplementary Material.

The detailed internal structure of the resistive layer is not
resolved by this exterior solution.  We therefore retain only the
minimal local assumptions required to obtain a controlled result.
In addition to the common reversal
\eqref{eq:main_common_reversal}, we assume that the reversal is
locally reflection symmetric in the field-line label $\Xi$,
\begin{equation}
 \mathcal Q''(0)=0 . 
 \label{eq:main_q_symmetric}
\end{equation}
The plasma-dependent term
${\mathfrak A}_{\rm pl}$ is likewise not fixed by force-free
electrodynamics. As shown in the Supplementary Material, it vanishes identically when the zeroth-order flow
is ideal. Hence any change of the inclination angle $\chi(t)$ must come from non-ideal effects.  
In the
following we set ${\mathfrak A}_{\rm pl}$ to zero and isolate the magnetic-diffusion
contribution.  Thus the result below describes the controlled
resistive mechanism rather than a model of unresolved current-sheet
plasma dynamics. With these assumptions, we find \begin{equation}
 {\mathfrak R}_{\rm diff}
 =
 -\Gamma_\chi(r)\chi ,
 \qquad
 \Gamma_\chi(r)
 =
 \frac{\eta_{\rm m}a_\star^2}{M^2}\,
 {\cal G}(r/M),
 \label{eq:main_general_rate}
\end{equation}
where the explicit function ${\cal G}$ and the cancellations leading
to Eq.~\eqref{eq:main_general_rate} are given in the Supplementary Material.
At the future horizon, we obtain \eqref{eq:mainresult} and hence
\begin{equation}
 \chi(t)\propto
 \exp(-\Gamma_\chi t).
 \label{eq:main_alignment_law}
\end{equation}
The mechanism therefore predicts two independent signatures:
the late-time alignment rate is linear in resistivity and quadratic
in black-hole spin. The late-time quadratic spin dependence of $ \Gamma_\chi$ was established numerically in Ref.~\cite{Selvi_2024}.

\section{Striped and warped current sheet}

Equation~\eqref{eq:main_BZ_euler} also fixes the geometry of a
consistent inclined split monopole and thereby provides the initial
condition for our simulations.  Reversing the smooth BZ field across
a surface $\Xi=0$ gives
\begin{equation}
 \overline F=\operatorname{sgn}(\Xi)F .
\end{equation}
The homogeneous Maxwell equation requires
$\dd\Xi\wedge F=0$ on the sheet.  Since the BZ field has the
Euler-potential form \eqref{eq:main_BZ_euler}, this condition is
satisfied by any sheet $\Xi=\Xi(\psi,\Phi)$
\cite{GrallaJacobson2014}. To describe an inclination $\chi$, define a magnetic colatitude
$\vartheta$ through
\begin{equation}
 \cos\vartheta=1-\frac{\psi}{\psi_0}.
 \label{eq:main_vartheta}
\end{equation}
The inclined split is then
\begin{equation}
 \Xi_{\rm BZ}
 =
 \cos\chi\cos\vartheta
 +\sin\chi\sin\vartheta\cos\Phi .
 \label{eq:main_exact_sheet}
\end{equation}
\begin{figure}
 \centering \includegraphics[width=1\linewidth]
{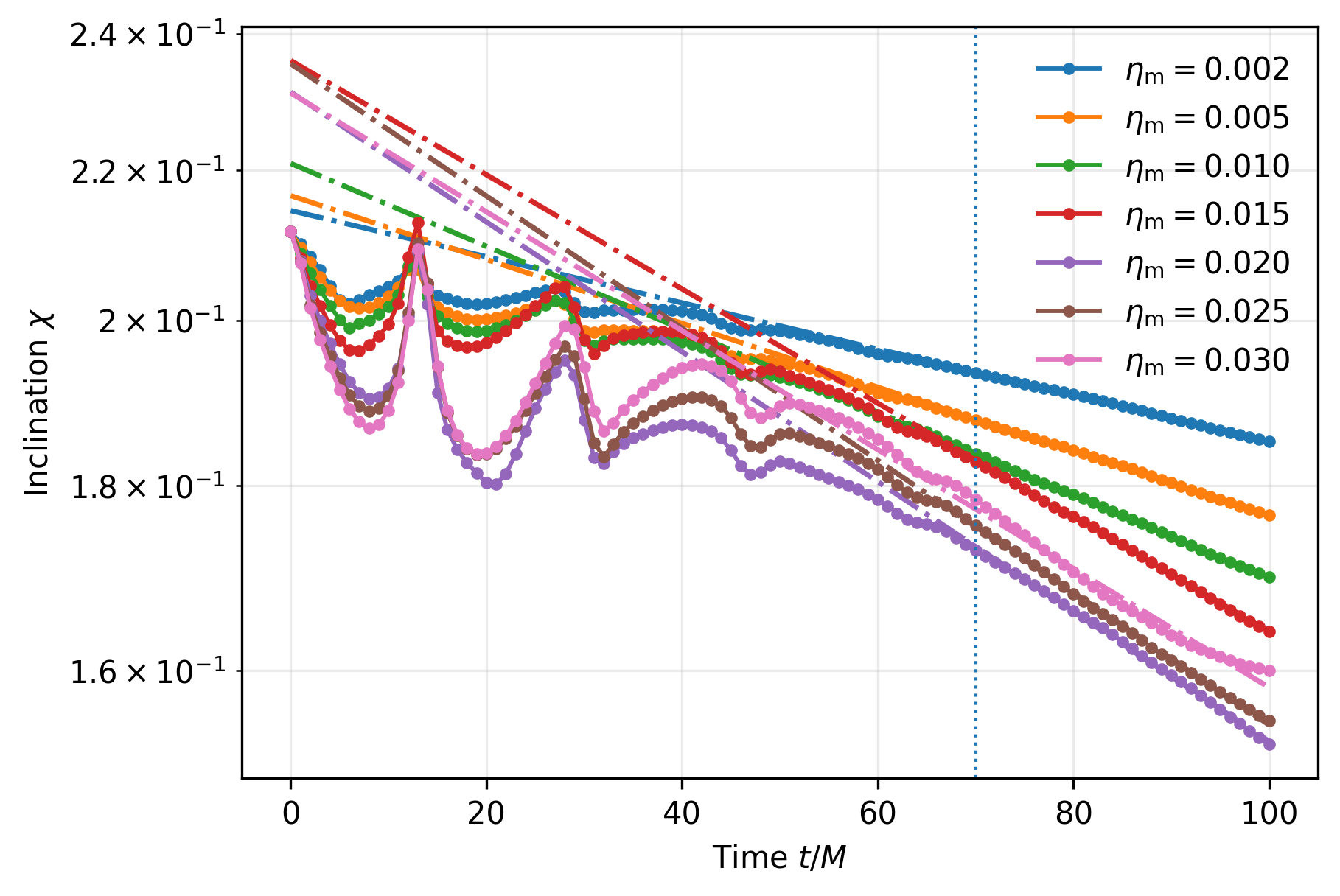}
 \caption{\label{fig:alignment_eta}
 Resistivity dependence of the alignment for $a_\star=0.6$. Solid curves show the numerical inclinations and dash-dotted curves an exponential fit starting from the dotted vertical line. The alignment rate corresponding to the exponential fit is plotted in Fig.~\ref{fig:alignment_eta111}.}
\end{figure}
The radial dependence of $\Phi$ produces the familiar striped
Bogovalov structure.  In addition, the $O(a_\star^2)$ deformation of
the BZ flux surfaces produces a radial warp.  Expanding
Eq.~\eqref{eq:main_vartheta} gives
\begin{equation}
 \vartheta
 =
 \theta
 +a_\star^2\Rhat(r/M)\sin\theta\cos\theta
 +O(a_\star^4),
\end{equation}
so that for $\chi\ll1$
\begin{equation}
 \theta_\Sigma
 =
 \frac{\pi}{2}
 +\chi
 \left[1+a_\star^2\Rhat(r/M)\right]\cos\Phi
 +O(\chi^2,a_\star^4) .
 \label{eq:main_warped_sheet}
\end{equation}
The angular location of the sheet used for the numerical initial condition is equivalently specified by
\begin{equation}
 S
 =
 \theta-\frac{\pi}{2}
 -\chi
 \left[1+a_\star^2\Rhat(r/M)\right]\cos\Phi .
 \label{eq:main_initial_condition}
\end{equation}
Here $S=0$ is only a geometric description of the reversal surface; the smooth analytic magnetic field is constructed with $\mathcal Q(\Xi_{\rm BZ})$, not with a profile of $S$. The dependence of the sheet on $\Phi$ gives rise to a spatially and temporally striped current sheet \cite{GrallaJacobson2014,Bogovalov:1999vg}. The $\mathcal{O} ( a_\star^2 ) $ correction induces additional warping of this current sheet and is a signature effect of a black-hole magnetosphere.  

\section{Numerical simulations}

We test Eq.~\eqref{eq:main_alignment_law} with three-dimensional resistive-GRMHD simulations using BHAC \cite{Porth_2017,Olivares_2019} and its resistive implementation \cite{Ripperda_2019,Palenzuela_2009}. The domain extends to $r=40M$ with inner radius inside the horizon and a base resolution of $100^3$ cells. The initial current-sheet location is given by Eq.~\eqref{eq:main_initial_condition}, with $\chi_0=0.2$, and a constant scalar resistivity $\eta_{\rm m}$. The fluid velocity is initialized with the ideal-MHD drift $
v^i=\frac{\epsilon^{ijk}E_jB_k}{B^2}$. The magnetic field is normalized to an initial maximum magnetization $b^2/\rho=10^2$. During the evolution we impose
\[
\rho_{\rm floor}
=
\max\!\left[\left(\frac{5M}{r}\right)^4,\frac{b^2}{10^3}\right],
\qquad
p_{\rm floor}=10^{-2}\rho_{\rm floor},
\]
thereby limiting the magnetization to $b^2/\rho\leq10^3$. Static mesh refinement is concentrated around the current sheet, with the highest refinement region wide enough to contain the full tilted sheet. At each time we locate the magnetic reversal on the horizon and extract its $m=1$ azimuthal Fourier amplitude, which measures the geometric tilt $\chi_{\rm num}(t)$.
\begin{figure}
 \centering \includegraphics[width=1\linewidth]
{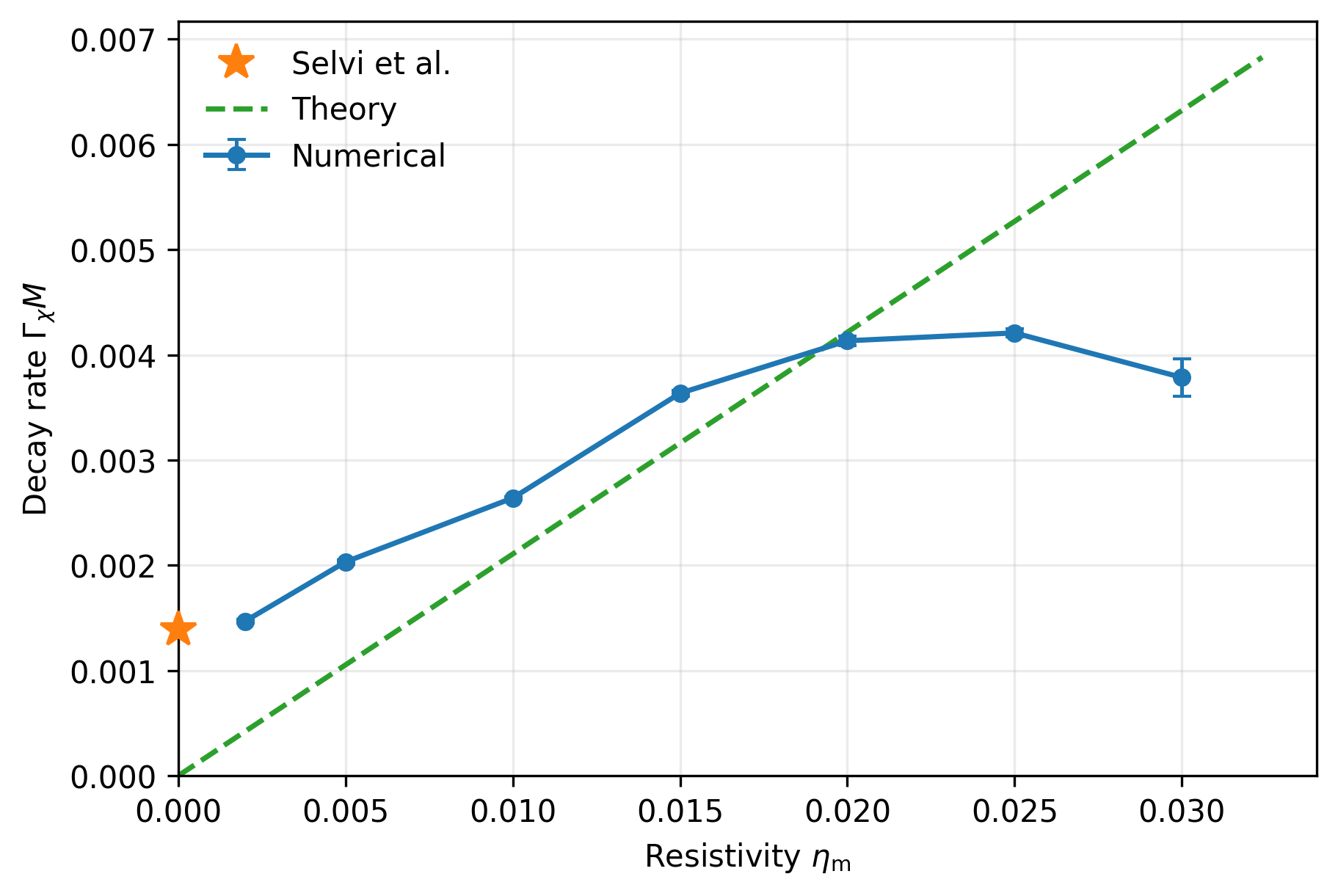} \caption{\label{fig:alignment_eta111} Alignment rate $\Gamma_\chi$ as a function of the explicit resistivity $\eta_{\rm m}$ for fixed black-hole spin $a_*=0.6$. 
The blue points show the decay rates obtained from late-time exponential fits to the numerical simulations, while the dashed line corresponds to Eq.~\eqref{eq:mainresult}. The star at $\eta_{\rm m}=0$ denotes the alignment rate inferred from the high-resolution ideal-GRMHD simulations of Ref.~\cite{Selvi_2024}, using their fit
$\tau_\chi/M \simeq 20(\Omega_H M)^{-2}$ and $\Gamma_\chi=1/\tau_\chi$ evaluated at $a_*=0.6$.}
\end{figure}
\begin{figure}
 \centering
 \includegraphics[width=1\linewidth]
 {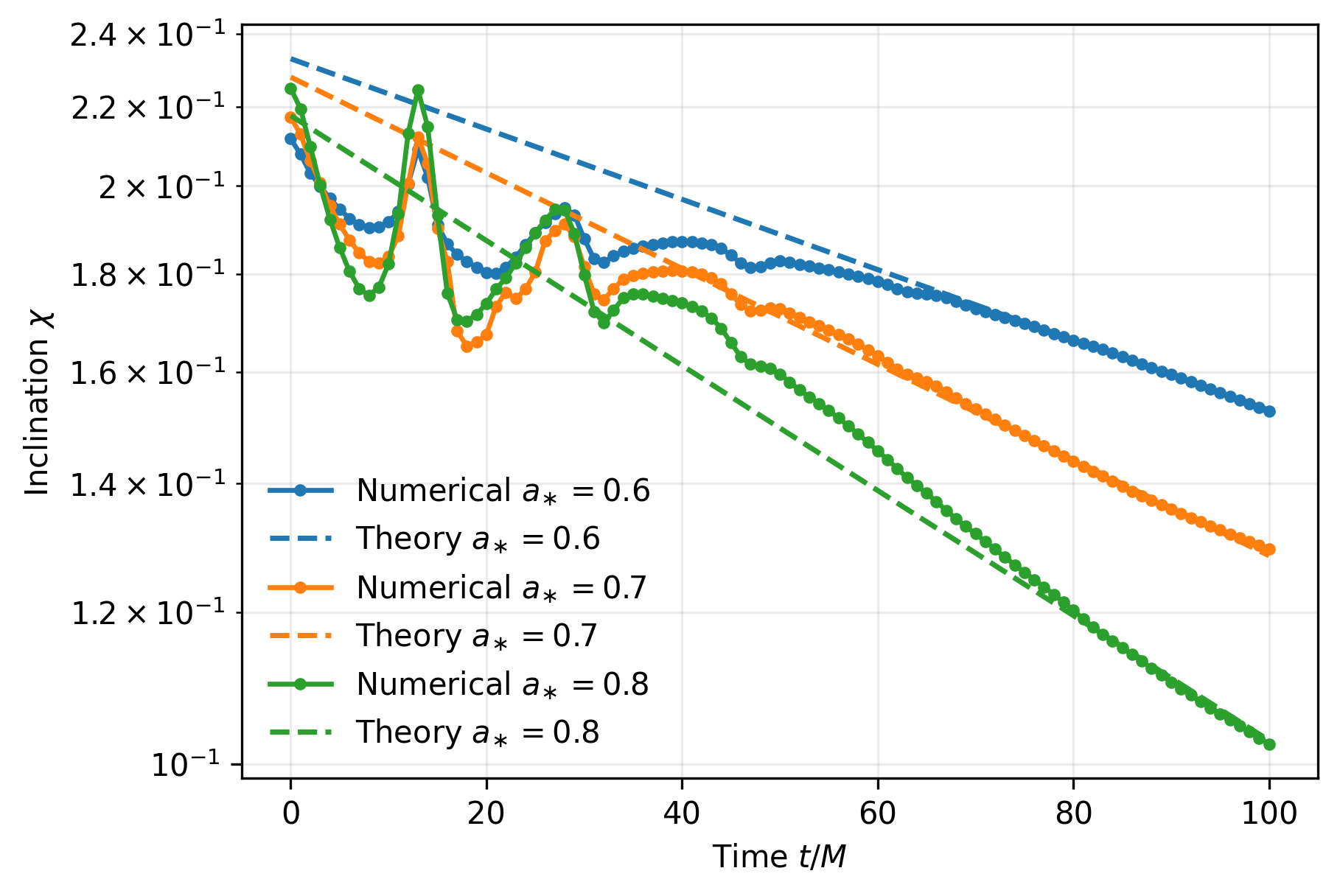}
 \caption{\label{fig:alignment_spin}
 Spin dependence of resistive alignment for $\eta_{\rm m} =0.02$. Solid curves show the
 numerical inclination for $a_\star=0.6$, $0.7$, and $0.8$; dashed
 curves show the theoretical prediction of Eq.~\eqref{eq:mainresult}, with coefficient $0.58544$. An additive offset in $\log\chi$ is fitted for the theoretical lines starting from $t = 75$.}
\end{figure}

\begin{figure}
    \centering    \includegraphics[width=0.95\linewidth]{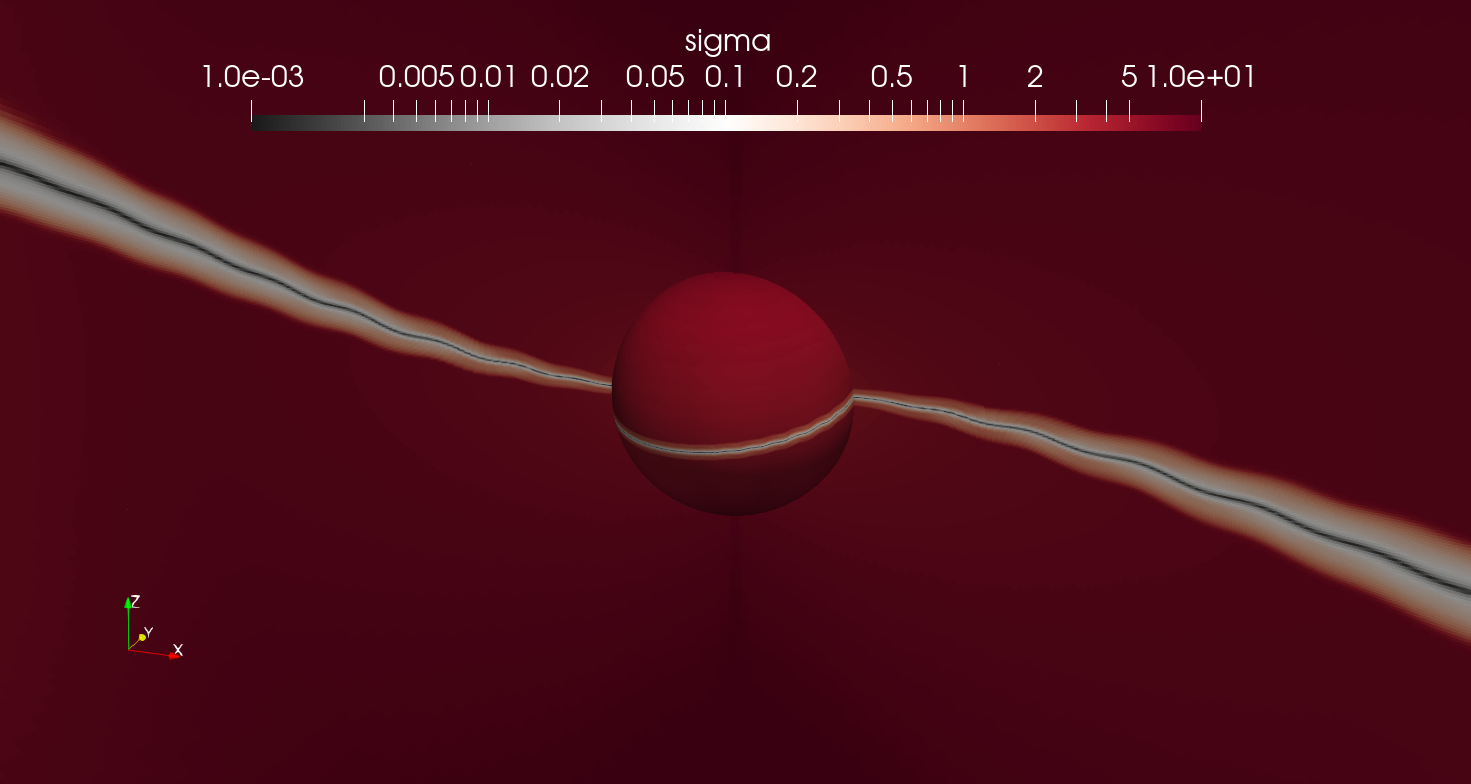}
    \caption{Snapshot of the magnetization $\sigma=b^2/\rho$ in a meridional slice of the black-hole magnetosphere at $t/ M = 96$ for $a_{\star } =0.6$ and $\eta_{\rm m  }  = 0.01 $. The low-magnetization region traces the current sheet, which exhibits low-amplitude wave-like distortions suggestive of the onset of a kink mode of the current sheet.}
    \label{fig:placeholder11}
\end{figure}

Fig.~\ref{fig:alignment_eta} shows $\chi_{\rm num}(t)$ at fixed $a_\star=0.6$ for different resistivities, with dash-dotted lines corresponding to late-time exponential fits. After an initial transient, the inclination settles onto a smooth monotonic decay. The fitted decay rates can be compared directly with the linear dependence predicted by Eq.~\eqref{eq:mainresult}. In Fig.~\ref{fig:alignment_eta111}, the fitted alignment rates are plotted against the theoretical prediction as a function of resistivity. At small resistivity, the measured rates exceed the magnetic-diffusion prediction. In particular, while the theory predicts $\Gamma_{\chi}\rightarrow0$ as $\eta_{\rm m}\rightarrow0$, the measured alignment rate instead approaches a finite plateau, signaling the onset of an MHD instability. This behavior is consistent with Ref.~\cite{Selvi_2024}, where alignment was measured in high-resolution ideal-GRMHD simulations and found to be insensitive to resolution. Moreover, the folds and warps of the current sheet observed in Ref.~\cite{Selvi_2024} also start to form in our low-resistivity simulations (Fig.~\ref{fig:placeholder11}), suggesting that MHD instabilities may play a role in the alignment process at low resistivity. The plateauing is not due to numerical resistivity, as the flux decay analysis in the Supplementary Material shows that numerical resistivity is too small to explain this effect. At large resistivity, the alignment rate in our simulations drops below the linear prediction. We attribute this to a breakdown of the force-free exterior approximation, as strong resistive diffusion modifies the magnetosphere away from the stationary Blandford–Znajek background assumed in the analytic calculation. 

Figure~\ref{fig:alignment_spin} compares the spin dependence at fixed resistivity $\eta_{\rm m}/M=0.020$ with the quadratic spin dependence predicted by Eq.~\eqref{eq:mainresult}. The theoretical rates are prescribed, while only the normalization of each exponential is fitted. 

\section{Discussion}
We have derived a prediction for the alignment of an inclined black-hole magnetosphere by combining the evolution equation for the magnetic-field reversal surface with the slow-spin Blandford--Znajek exterior solution. At leading order, the alignment rate scales as $  \Gamma_\chi \propto \eta_{\rm m} a_*^2 $. Our resistive GRMHD simulations provide a comparison with this prediction across a range of resistivities. At sufficiently large resistivity, the measured rate falls below the linear prediction, suggesting that resistive evolution of the magnetosphere can no longer be treated as a perturbation of the stationary Blandford–Znajek exterior assumed in the analytic calculation. At smaller resistivity, the alignment rate no longer decreases linearly with $\eta_{\rm m}$ but instead approaches an approximately constant value. This is consistent with the high-resolution ideal-GRMHD results of Ref.~\cite{Selvi_2024}, which also show finite-rate alignment in the absence of explicit resistivity and independent of numerical resolution. 
A possible explanation for the alignment even in the 'ideal' case is the onset of a current sheet instability similar to the kink mode discussed in the context of the Earth magnetotail current sheet \cite{ErkaevSemenovEtAl2007}.

Most importantly, our results reveal the physical mechanism behind the alignment process: the non-monopolar nature of the Blandford--Znajek flux function at second order in spin leads to asymmetric upstream conditions across an inclined current sheet and a nonlinear relation between the magnetic sheet label and polar angle. Both contributions enter the magnetic diffusion of a smooth, constraint-compatible reversal and drive the sheet toward alignment. \textcolor{black}{A possible connection between the spin-dependent nonuniformity of the magnetic field and current-sheet alignment was previously suggested in Ref.~\cite{Selvi_2024}.} 

In astrophysical black-hole magnetospheres the microscopic resistivity is expected to be very small, giving rise to kink- and tearing instabilities accompanied by the ejection of plasmoids \cite{Loureiro_2007,Bhattacharjee_2009,PhysRevLett.103.105004}. It would be interesting to explicitly study the role of instabilities in tilted current sheet dynamics and correlate their appearance with the measured alignment rate.

Additionally, astrophysical magnetospheres are likely in the collisionless regime, where reconnection rates can be enhanced by up to an order of magnitude \cite{PhysRevLett.127.055101,ripperda2026magneticreconnection01rate,Moran_2025,most2026nonlinearlycausalgeneralrelativistictwofluid}. It would therefore be interesting to investigate how the alignment rate is modified in models that properly capture the collisionless plasma dynamics.

The present calculation also relies on the slow-spin Blandford--Znajek expansion, so higher-order corrections may become important at large spin. It would therefore be interesting to apply the method developed here to other black-hole magnetosphere configurations, in particular the fast-spin Blandford--Znajek solution \cite{Nathanail2014} and the hyperbolic thin-disk magnetosphere \cite{PhysRevD.93.044038}.

\section{Acknowledgements}
RL is supported by the NWA ORC programme Emergence at All Scales. 
\textcolor{black}{OP acknowledges discussions with Sebastiaan Selvi, Bart Ripperda, Lorenzo Sironi and Jonathan Granot.}
The numerical simulations of this work were carried out on the Dutch national supercomputer Snellius, using the national e-infrastructure provided by SURF with support from the Dutch Research Council (NWO) through grant no. EINF-20236.

\section{Use of Large Language Models}
RL used ChatGPT (OpenAI, GPT-6 Astra) to assist with algebraic checks, code debugging, and the presentation of parts of the manuscript. All AI-assisted calculations and suggestions were independently checked by the authors, who take full responsibility for the content of the manuscript.

\appendix

\onecolumngrid

\newpage

\section*{Supplementary Material}
\label{eq:appendix}
\appendix
\renewcommand{\appendixname}{Section}
In this Supplementary Material we detail the derivations used in this work. Section~\ref{app:derivation} derives the main result of the paper, which is the analytical prediction of the resistive alignment rate. Section~\ref{app:warped_sheet} discusses the consistent splitting of the Blandford-Znajek monopole solution for an inclined current sheet. Section~\ref{eq:fluxdecay} provides analysis of the flux-decay in the performed simulations and gives an estimate of the residual numerical resistivity. Section~\ref{eq:convergence} shows the robustness of our results against numerical resolution.

\section{Derivation of alignment rate}
\label{app:derivation}
In this section we derive the alignment time of an inclined black-hole magnetosphere by combining the slow-spin Blandford-Znajek monopole solution with the equations of resistive relativistic magnetohydrodynamics that allow for an expression of the irreversible evolution of the inclination angle $\chi$.  
\subsection{The $3+1$ split}
We use geometric units $G=c=1$ and metric signature $(-,+,+,+)$. We foliate spacetime by surfaces of constant coordinate time, which gives \cite{Arnowitt2008}
\begin{equation}
 \dd s^2
 =-\Nlapse^2\dd t^2
 +\gamma_{ij}(\dd x^i+\beta^i\dd t)
                (\dd x^j+\beta^j\dd t).
 \label{eq:3plus1_metric}
\end{equation}
Here $\Nlapse$ is the lapse, $\beta^i$ is the shift, and
$\gamma_{ij}$ is the spatial metric.  The future-directed unit normal to
the slices is
\begin{equation}
 n_\mu=(-\Nlapse,0,0,0),
 \qquad
 n^\mu=\frac{1}{\Nlapse}(1,-\beta^i),
 \label{eq:Eulerian_observer}
\end{equation}
and defines the Eulerian observer.  The four- and three-dimensional
volume factors obey
\begin{equation}
 \sqrt{-g}=\Nlapse\sqrt\gamma,
 \qquad
 \gamma=\det\gamma_{ij}.
\end{equation}
The electric and magnetic fields measured by the Eulerian observer are
\begin{equation}
 E_\mu=F_{\mu\nu}n^\nu,
 \qquad
 B^\mu={{}^\star F}^{\mu\nu}n_\nu,
 \qquad
 E_\mu n^\mu=B_\mu n^\mu=0.
 \label{eq:Eulerian_fields}
\end{equation}
The four-current is decomposed as
\begin{equation}
 J^\mu=\rho_{\rm e}n^\mu+j^\mu,
 \qquad
 \rho_{\rm e}=-n_\mu J^\mu,
 \qquad
 n_\mu j^\mu=0,
 \label{eq:current_split}
\end{equation}
where $\rho_{\rm e}$ and $j^i$ are the charge density and spatial
current measured by the Eulerian observer. The symbol $D_i$ denotes the Levi--Civita covariant derivative of the
spatial metric $\gamma_{ij}$.  Thus
\begin{equation}
 D_i f=\partial_i f,
 \qquad
 D_iV^j=\partial_iV^j+{}^{(3)}\Gamma^j{}_{ik}V^k,
 \qquad
 D_iV_j=\partial_iV_j-{}^{(3)}\Gamma^k{}_{ij}V_k,
 \label{eq:spatial_covariant_derivative}
\end{equation}
with
\begin{equation}
 {}^{(3)}\Gamma^i{}_{jk}
 =\frac12\gamma^{i\ell}
 \left(
 \partial_j\gamma_{\ell k}
 +\partial_k\gamma_{\ell j}
 -\partial_\ell\gamma_{jk}
 \right),
 \qquad
 D_i\gamma_{jk}=0.
\end{equation}
In particular,
\begin{equation}
 D_iV^i=\frac{1}{\sqrt\gamma}\partial_i(\sqrt\gamma V^i),
 \qquad
 (\operatorname{curl}V)^i=\epsilon^{ijk}D_jV_k,
 \label{eq:spatial_div_curl}
\end{equation}
where $\epsilon^{ijk}$ is the spatial Levi--Civita tensor, normalized by
$\epsilon^{123}=1/\sqrt\gamma$.

\subsection{Maxwell equations}
The electromagnetic field is the two-form $F_{\mu\nu}$.  With the sign convention for the electric field used below, Maxwell's equations are written as
\begin{equation}
 \nabla_\nu  {{}^\star F}^{\mu\nu}=0,
 \qquad
 \nabla_\nu F^{\mu\nu} = J^\mu  , 
 \label{eq:covariant_Maxwell}
\end{equation}
where 
\begin{equation}
 {{}^\star F}^{\mu\nu}
 =\frac12\epsilon^{\mu\nu\rho\sigma}F_{\rho\sigma}.
\end{equation}
The first equation contains Faraday's law and the no-monopole
constraint.  The second contains Amp\`ere's law and Gauss's law for the
electric field. The homogeneous equation in Eq.~\eqref{eq:covariant_Maxwell} gives the
magnetic constraint
\begin{equation}
 {
 \partial_i(\sqrt\gamma B^i)=0
 }
 \label{eq:divB_3plus1}
\end{equation}
and the exact $3+1$ Faraday equation \cite{DelZanna2007}
\begin{equation}
 \partial_t(\sqrt\gamma B^i)
 +\partial_j\!\left[
 \sqrt\gamma\left(
 \Nlapse\epsilon^{ijk}E_k
 +\beta^iB^j-\beta^jB^i
 \right)
 \right]
 =0.
 \label{eq:Faraday_3plus1_exact}
\end{equation}
The inhomogeneous equation in Eq.~\eqref{eq:covariant_Maxwell} gives
Gauss's law and Amp\`ere's law in the standard resistive-GRMHD
$3+1$ form~\cite{PhysRevD.88.044020}.
\begin{equation}
 {
 D_iE^i=\rho_{\rm e}
 }
 \label{eq:GaussE_3plus1}
\end{equation}
and the exact conservative Amp\`ere equation
\begin{equation}
 {
 \begin{aligned}
 \partial_t(\sqrt\gamma E^i)
 &+\partial_j\!\left\{
 \sqrt\gamma\left[
 \beta^iE^j-\beta^jE^i
 -\Nlapse\epsilon^{ijk}B_k
 \right]
 \right\}
 =-\sqrt\gamma
 \left(\Nlapse j^i-\rho_{\rm e}\beta^i\right).
 \end{aligned}
 }
 \label{eq:Ampere_3plus1_exact}
\end{equation}
It is useful to rewrite this as an explicit expression for the
Eulerian current. Define the extrinsic curvature by
\begin{equation}
 K_{ij}=-\frac12\mathcal L_n\gamma_{ij},
 \qquad
 K=\gamma^{ij}K_{ij},
\end{equation}
and the spatial Lie derivative
\begin{equation}
 (\mathcal L_\beta E)^i
 =\beta^jD_jE^i-E^jD_j\beta^i.
\end{equation}
Using Eq.~\eqref{eq:GaussE_3plus1} and
$\partial_t\sqrt\gamma=\sqrt\gamma(-\Nlapse K+D_i\beta^i)$,
Eq.~\eqref{eq:Ampere_3plus1_exact} becomes
\begin{equation}
 {
 j^i
 =\frac{1}{\Nlapse}\epsilon^{ijk}D_j(\Nlapse B_k)
 +K E^i
 -\frac{1}{\Nlapse}
 \left(\partial_t-\mathcal L_\beta\right)E^i.
 }
 \label{eq:Ampere_current_exact}
\end{equation}
Finally, \eqref{eq:Faraday_3plus1_exact} can also be written as 
\begin{equation}  \label{eq:faraday11111}
 \partial_t   B^i  - \mathcal{L}_{\beta} B^i   
 +  D_j \left(
 \Nlapse\epsilon^{ijk}E_k  \right)  =  \Nlapse K B^i .    \end{equation}

\subsection{Resistive closure and the magnetic-diffusion limit}

We adopt the simplest scalar relativistic Ohm law,
\begin{equation}
 J^\mu=\rho_0u^\mu+\sigma e^\mu,
 \qquad
 e^\mu=F^{\mu\nu}u_\nu,
 \label{eq:covariant_Ohm}
\end{equation}
where $u^\mu$ is the fluid four-velocity, $\rho_0=-u_\mu J^\mu$
is the charge density measured in the fluid frame, and $\sigma$ is the
scalar conductivity.  We decompose
\begin{equation}
 u^\mu=\Gamma(n^\mu+v^\mu),
 \qquad
 n_\mu v^\mu=0,
 \qquad
 \Gamma=(1-v^2)^{-1/2},
\end{equation}
and decompose the four-current with respect to the Eulerian observer as
\begin{equation}
 J^\mu=q\,n^\mu+j^\mu,
 \qquad
 q\equiv-n_\mu J^\mu,
 \qquad
 n_\mu j^\mu=0.
 \label{eq:Eulerian_current_decomposition}
\end{equation}
Here $q$ and $j^\mu$ are respectively the Eulerian charge density and
spatial current. The exact spatial projection of Eq.~\eqref{eq:covariant_Ohm} is then
\begin{equation}
 j^i
 =
 qv^i
 +\sigma \Gamma
 \left[
 E^i+\epsilon^{ijk}v_jB_k
 -(E^jv_j)v^i
 \right].
 \label{eq:Eulerian_Ohm}
\end{equation}
The comoving and Eulerian charge densities are related by
\begin{equation}
 q
 =
 \Gamma\rho_0+\sigma \Gamma E^iv_i,
 \label{eq:charge_density_relation}
\end{equation}
but we will work with the Eulerian variables in
Eq.~\eqref{eq:Eulerian_Ohm}, so that all non-ideal contributions remain
explicit. Define the ideal electric field and scalar magnetic resistivity by
\begin{equation}
 E_{\rm id}^i
 \equiv
 -\epsilon^{ijk}v_jB_k,
 \qquad
 \eta_{\rm m}\equiv\frac{1}{\sigma}.
 \label{eq:Eid_eta_definitions}
\end{equation}
Equation~\eqref{eq:Eulerian_Ohm} may then be rearranged exactly as
\begin{equation}
 E^i
 =
 E_{\rm id}^i
 +\frac{\eta_{\rm m}}{\Gamma}
 \left(j^i-qv^i\right)
 +(E^jv_j)v^i.
 \label{eq:E_resistive_before_reductions}
\end{equation}
Our aim is to isolate the magnetic-diffusion contribution to the secular
motion of the current sheet without attempting to solve for the full plasma
dynamics.  We therefore make a sequence of explicit reductions in the non-ideal sector. First, following the standard long-timescale reduction
\cite{Geppert_2000,Rezzolla_2001}, we neglect the electric-field
contribution to the Eulerian current in
Eq.~\eqref{eq:Ampere_current_exact}.  In the present $3+1$ notation this amounts to the ordering
\begin{equation}
 \left|
 K E
 -\frac{1}{\Nlapse}
 \left(\partial_t-\mathcal L_\beta\right)E
 \right|
 \ll
 \left|
 \frac{1}{\Nlapse}\operatorname{curl}(\Nlapse B)
 \right|.
 \label{eq:low_frequency_ordering}
\end{equation}
Thus we drop
\begin{equation}
 K E^i
 -\frac{1}{\Nlapse}
 \left(\partial_t-\mathcal L_\beta\right)E^i
 \label{eq:full_displacement_piece}
\end{equation}
relative to the magnetic-curl contribution.  Defining
\begin{equation}
 \mathcal J_B^i
 \equiv
 \frac{1}{\Nlapse}
 \epsilon^{ijk}D_j(\Nlapse B_k),
 \label{eq:JB_definition}
\end{equation}
the reduced Amp\`ere equation is
\begin{equation}
 j^i\simeq\mathcal J_B^i.
 \label{eq:reduced_Ampere}
\end{equation}
Second, in the non-ideal part of Ohm's law we neglect the charge-advection
term
\begin{equation}
 -\frac{\eta_{\rm m}}{\Gamma}qv^i.
 \label{eq:discard_charge_advection}
\end{equation}
Third, we neglect the longitudinal electric-field contribution
\begin{equation}
 (E^jv_j)v^i.
 \label{eq:discard_longitudinal_E}
\end{equation}
Finally, because the stationary exterior magnetic solution does not determine the plasma dynamics, we adopt the simplest isotropic magnetic-diffusion closure by neglecting Lorentz-factor corrections to the non-ideal response, setting
\begin{equation}
 \Gamma\simeq1
\end{equation}
in the resistive term. Under these approximations,
Eq.~\eqref{eq:E_resistive_before_reductions} reduces to
\begin{equation}
 E^i
 \simeq
 E_{\rm id}^i
 +\frac{\eta_{\rm m}}{\Gamma}j^i
 \simeq
 E_{\rm id}^i+\eta_{\rm m}j^i,
 \label{eq:reduced_Ohm_intermediate}
\end{equation}
and using Eq.~\eqref{eq:reduced_Ampere},
\begin{equation}
 E^i
 \simeq
 E_{\rm id}^i+\eta_{\rm m}\mathcal J_B^i.
 \label{eq:reduced_Ohm}
\end{equation}
Substitution of \eqref{eq:reduced_Ampere} and  \eqref{eq:reduced_Ohm} into the exact Faraday equation of \eqref{eq:Faraday_3plus1_exact} gives the quasistatic
curved-space induction equation
\begin{equation}
 \begin{aligned}
 \partial_t(\sqrt\gamma B^i)
 +\partial_j\!\Bigg\{
 \sqrt\gamma\Big[&
 V^jB^i-V^iB^j
 +\eta_{\rm m}
 \big(
 D^i(\Nlapse B^j)-D^j(\Nlapse B^i)
 \big)
 \Big]\Bigg\}=0,
 \end{aligned} \label{eq:resistive_induction_quasistatic}
\end{equation}
where
\begin{equation}
 V^i=\Nlapse v^i-\beta^i
 \label{eq:transport_velocity}
\end{equation}
is the coordinate transport velocity.

\subsection{From the vector induction equation to a signed sheet marker}
We now wish to turn the induction equation of
Eq.~\eqref{eq:resistive_induction_quasistatic} into an expression for the
evolution of the current sheet.  From this point onward we use spherical
spatial coordinates
\begin{equation}
 x^i=(r,\theta,\phi),
\end{equation}
adapted to the black-hole spin axis: $r$ is the radial coordinate,
$\theta$ is the polar angle measured from the spin axis, and $\phi$ is the
azimuthal angle.  The precise Kerr geometry and horizon-regular coordinate
realization will be specified later.  For the present kinematic construction,
only this spherical labeling of the spatial directions is required. To follow the motion of the current sheet, we need a signed scalar whose zero
defines its reversal surface $\Sigma$.  For the black-hole magnetosphere that
we consider in this work, which corresponds to a split-monopole geometry, the
sheet is a deformation of the equatorial surface $\theta=\pi/2$ and the
magnetic polarity reverses in the radial component.  The natural choice is
therefore
\begin{equation}
 {\mathcal B\equiv B^r.}
\end{equation}
$B^r$ changes sign across the split and the sheet is tracked by
$\mathcal B=0$. Taking the radial component, $i=r$, in
Eq.~\eqref{eq:resistive_induction_quasistatic}, the $j=r$ flux vanishes
identically by antisymmetry.  The complete radial component is therefore
\begin{equation}
 {
 \begin{aligned}
 \partial_t(\sqrt\gamma\,\mathcal B)
 +\partial_a\!\Big\{\sqrt\gamma\big[&
 V^a\mathcal B-V^rB^a +\eta_{\rm m}
 \big(D^r(\Nlapse B^a)-D^a(\Nlapse\mathcal B)\big)
 \big]\Big\}=0,
 \end{aligned}}
 \label{eq:r_component_full}
\end{equation}
where $a\in\{\theta,\phi\}$.  On the reversal surface $\Sigma$,
$\mathcal B=0$, so that Eq.~\eqref{eq:r_component_full} becomes
\begin{equation}
 {
 \begin{aligned}
 (\partial_t+V^a\partial_a)\mathcal B
 =\frac1{\sqrt\gamma}\partial_a\!\Big\{\sqrt\gamma\big[&
 V^rB^a +\eta_{\rm m}
 \big(D^a(\Nlapse\mathcal B)-D^r(\Nlapse B^a)\big)
 \big]\Big\}
 \end{aligned}}
 \quad\hbox{on }\Sigma .
 \label{eq:scalar_induction_full_at_zero}
\end{equation}
The resistive term in Eq.~\eqref{eq:scalar_induction_full_at_zero}
will be kept in the form in which it arises from the reduced Amp\`ere law.
Define the full magnetic-diffusion contribution
\begin{equation}
 \mathcal C_{\rm diff}
 \equiv \frac1{\sqrt\gamma}\partial_a\!\left\{
 \sqrt\gamma\eta_{\rm m}
 \left[D^a(\Nlapse\mathcal B)-D^r(\Nlapse B^a)\right]
 \right\}, \label{eq:Cdiff_general_definition}
\end{equation}
Then Eq.~\eqref{eq:scalar_induction_full_at_zero} is simply
\begin{equation}
 {
 (\partial_t+V^a\partial_a)\mathcal B
 =
 \frac1{\sqrt\gamma}\partial_a(\sqrt\gamma V^rB^a)
 +\mathcal C_{\rm diff}
 }
 \quad\hbox{on }\Sigma .
\label{eq:scalar_induction_reduced_at_zero}
\end{equation}

\subsection{The reversal surface and corotating phase}

The construction used here is analogous to the
kinematic argument used to follow magnetic X-lines and null points in
Refs.~\cite{murphymurphy,parnell,10.1063/1.860718}: differentiating the defining
condition that the relevant magnetic field vanishes gives the velocity of its
zero.  Those works track an isolated magnetic null (or, under translational
symmetry, an X-line), whereas here the zero of the reversing component
$\mathcal B=B^r$ defines an extended current-sheet surface.  Locally, this
surface may be written as
\begin{equation}
 \theta=\theta_\Sigma(t,r,\phi),
\end{equation}
with
\begin{equation}
 \mathcal B(t,r,\theta_\Sigma,\phi)=0,
 \qquad
 \left.\partial_\theta\mathcal B\right|_\Sigma\neq0.
 \label{eq:reversal_surface}
\end{equation}
A small inclination of a current sheet that would otherwise coincide with the black-hole equatorial plane, i.e. the plane orthogonal to the spin axis, produces a single-wave azimuthal distortion around the hole. In a Fourier decomposition in the azimuthal angle \(\phi\), this is the \(m=1\) harmonic: the sheet is displaced once above and once below the equatorial plane during one full \(2\pi\) rotation in \(\phi\). The current sheet is forced to corotate in order to obey the no-monopole constraint \cite{GrallaJacobson2014,Bogovalov:1999vg}, which gives it a striped profile. As will be shown, the corotating phase of this harmonic is given by
\begin{equation}
 \Phi=\phi-\OmF[t-\mathcal M(r)]-\Phi_0,
 \label{eq:general_outgoing_phase}
\end{equation}
where $\OmF$ is the field-line angular velocity and $\mathcal M(r)$ allows
for a radial phase lag or winding of the sheet.  Its explicit form will only
be needed once the Blandford--Znajek background is specified. At fixed $r$ and fixed phase $\Phi$, the corresponding time derivative is
\begin{equation}
 {
 \DtPhi=\left.\partial_t\right|_{r,\Phi}
 =\partial_t+\OmF\partial_\phi.}
 \label{eq:corotating_derivative}
\end{equation}
The function $\mathcal R(r)$ affects radial derivatives but not this
fixed-radius time derivative. Differentiating the reversal condition gives
\begin{equation}
 0=\left[
 \DtPhi\mathcal B
 +(\DtPhi\theta_\Sigma)\partial_\theta\mathcal B
 \right]_\Sigma.
 \label{eq:surface_chain_rule}
\end{equation}
Using Eq.~\eqref{eq:scalar_induction_reduced_at_zero} yields the surface-velocity relation
\begin{equation}
 {
 \begin{aligned}
 \DtPhi\theta_\Sigma={}&V^\theta
 +(V^\phi-\OmF)
 \frac{\partial_\phi\mathcal B}{\partial_\theta\mathcal B}
 -\frac{1}{\sqrt\gamma\,\partial_\theta\mathcal B}
 \partial_a(\sqrt\gamma V^rB^a)
 -\frac{\mathcal C_{\rm diff}}{\partial_\theta\mathcal B}
 \end{aligned}}
 \quad\hbox{on }\Sigma.
 \label{eq:full_surface_velocity}
\end{equation}
The first three terms are plasma advection and the fourth is the full
magnetic-diffusion contribution.

\subsection{Inner profile and the pre-projection dictionary}

At small inclination, we prescribe one inclination parameter in magnetic colatitude,
\begin{equation}
 \chi=\chi(t),
 \label{eq:alpha_time_only}
\end{equation}
and write the corresponding reversal surface as
\begin{equation}
 \theta_\Sigma(r,\Phi,t)=\frac{\pi}{2}+\chi(t)C(r)\cos\Phi
 +O(\chi^3,a_\star^4\chi),\qquad
 C(r)=1+a_\star^2\Rhat(r/M).
 \label{eq:small_alpha_sheet}
\end{equation}
The local horizon rate will be evaluated using this prescribed radial shape; we do not solve for a general amplitude $\chi(t,r)$. At fixed radius, $C(r)$ is time independent and does not affect a logarithmic decay rate.
We factorize the reversing field using the Euler-potential label of the sheet:
\begin{equation}
 \mathcal B=A_B(r,\theta)\mathcal Q(\Xi),\qquad
 \Xi=\cos\chi\cos\vartheta+\sin\chi\sin\vartheta\cos\Phi,
 \qquad \cos\vartheta=1-\frac{\psi}{\psi_0}.
 \label{eq:theta_inner_outer}
\end{equation}
Here $A_B>0$ is the smooth unsigned exterior amplitude of $B^r$, and $\mathcal Q$ describes the unresolved reversal. Its constant width normalization is absorbed into the function itself. We require
\begin{equation}
 \mathcal Q(0)=0,\qquad \mathcal Q'(0)\ne0,
 \label{eq:q_minimal_conditions}
\end{equation}
so that the zero is simple. For the coefficient calculated below we assume
\begin{equation}
 \mathcal Q''(0)=0.
 \label{eq:q_symmetric_assumption}
\end{equation}
The value of $\mathcal Q'(0)$ cancels from all derivative ratios, and no higher derivative of $\mathcal Q$ enters the diffusion operator evaluated at the zero.

There are two distinct sources of polar curvature. The exterior amplitude varies across an inclined sheet:
\begin{equation}
 \left.\partial_\theta\ln A_B\right|_\Sigma
 =\chi C\cos\Phi\left.\partial_\theta^2\ln A_B\right|_{\pi/2}
 +O(\chi^3,a_\star^4\chi).
 \label{eq:outer_envelope_asymmetry}
\end{equation}
In addition, $\Xi$ is nonlinear in $\theta$. Its second polar derivative cannot be dropped even when $\mathcal Q''(0)=0$. The following dictionary keeps both effects.

\subsection{Local derivative ratios on the reversal}

All derivatives are taken before evaluating at $\Xi=0$. Let us write $\Xi_c=\partial_c\Xi$ and $\Xi_{cd}=\partial_c\partial_d\Xi$. Then
\begin{equation}
 \left.\partial_\theta\mathcal B\right|_\Sigma
 =\left.A_B\mathcal Q'(0)\Xi_\theta\right|_\Sigma,
 \label{eq:first_theta_derivative}
\end{equation}
and define
\begin{equation}
 \mathcal R_c=\left.\frac{\partial_c\mathcal B}{\partial_\theta\mathcal B}\right|_\Sigma,
 \qquad
 \mathcal R_{cd}=\left.\frac{\partial_c\partial_d\mathcal B}{\partial_\theta\mathcal B}\right|_\Sigma.
 \label{eq:R_dictionary_definition}
\end{equation}
The chain rule gives
\begin{align}
 \mathcal R_c&=\left.\frac{\Xi_c}{\Xi_\theta}\right|_\Sigma,
 \label{eq:Xi_first_dictionary}\\
 \mathcal R_{cd}&=\left[
 \frac{\Xi_{cd}}{\Xi_\theta}
 +(\partial_c\ln A_B)\frac{\Xi_d}{\Xi_\theta}
 +(\partial_d\ln A_B)\frac{\Xi_c}{\Xi_\theta}
 +\frac{\mathcal Q''(0)}{\mathcal Q'(0)}\frac{\Xi_c\Xi_d}{\Xi_\theta}
 \right]_\Sigma.
 \label{eq:Xi_second_dictionary}
\end{align}
In particular,
\begin{equation}
 \mathcal R_{\theta\theta}
 =\left[2\partial_\theta\ln A_B+
 \frac{\Xi_{\theta\theta}}{\Xi_\theta}
 +\frac{\mathcal Q''(0)}{\mathcal Q'(0)}\Xi_\theta\right]_\Sigma  , 
 \label{eq:theta_derivative_ratio}
\end{equation}
where 
\begin{equation}
 \left.\frac{\Xi_{\theta\theta}}{\Xi_\theta}\right|_\Sigma
 =\left.\frac{\partial_\theta^2\vartheta}{\partial_\theta\vartheta}\right|_\Sigma
 =4a_\star^2\Rhat(x)\chi\cos\Phi
 +O(a_\star^4\chi,a_\star^2\chi^3).
 \label{eq:Xi_polar_curvature}
\end{equation}
Using Eq.~\eqref{eq:small_alpha_sheet} and $\mathcal Q''(0)=0$, the other ratios through linear order in $\chi$ and through the required spin order are
\begin{align}
 \mathcal R_\phi&=\chi C\sin\Phi+O(\chi^3,a_\star^4\chi),
 \label{eq:phi_theta_ratio}\\
 \mathcal R_{\phi\phi}&=\chi C\cos\Phi+O(\chi^3,a_\star^4\chi),
 \label{eq:phi_second_ratio}\\
 \mathcal R_r&=\chi CP\sin\Phi-\chi C'\cos\Phi+\cdots,
 \label{eq:r_theta_ratio}\\
 \mathcal R_{\theta r}&=\partial_r\ln A_B-C'
 +O(\chi^2/M,a_\star^4/M),
 \label{eq:mixed_theta_r_ratio}\\
 \mathcal R_{\theta\phi}&=O(\chi^2),
 \label{eq:mixed_theta_phi_ratio}\\
 \mathcal R_{\phi r}&=\chi C\left[P\cos\Phi+(\partial_r\ln A_B)\sin\Phi\right]+\cdots.
 \label{eq:mixed_radial_phi_ratio}
\end{align}
Here $x=r/M$, $C(r)=1+a_\star^2\Rhat(r/M)$, $C'=\mathrm d C/\mathrm dr$, primes on $\Rhat$ denote $x$ derivatives, and the ellipses denote terms beyond the retained spin and inclination orders. 
The radial phase gradient is
\begin{equation}
 P(r)=\partial_r\Phi.
 \label{eq:P_general_definition}
\end{equation}
The nonradial magnetic components share the same profile:
\begin{equation}
 B^r=\mathcal B,\qquad B^a=\lambda^a(r,\theta)\mathcal B,
 \qquad \lambda^a=B^a_{\rm ext}/B^r_{\rm ext}.
 \label{eq:common_reversal_profile}
\end{equation}
This is an assumption about the interior field, now made compatible with the magnetic constraint throughout the layer. The exterior ratios have the equatorial behavior
\begin{equation}
 \left.\lambda^\theta\right|_\Sigma
 =\ell(r)\chi C\cos\Phi+O(\chi^3,a_\star^4\chi),
 \qquad
 \left.\lambda^\phi\right|_\Sigma=\lambda^\phi_0(r)+O(\chi^2),
 \qquad
 \ell=\left.\partial_\theta\lambda^\theta\right|_{\pi/2}.
 \label{eq:lambda_expansion}
\end{equation}
For the derivatives of the nonradial components,
\begin{align}
 \left.\frac{\partial_cB^a}{\partial_\theta\mathcal B}\right|_\Sigma
 &=\lambda^a\mathcal R_c,
 \label{eq:Ba_first_derivative_dictionary}\\
 \left.\frac{\partial_c\partial_dB^a}{\partial_\theta\mathcal B}\right|_\Sigma
 &=\lambda^a\mathcal R_{cd}+(\partial_c\lambda^a)\mathcal R_d
 +(\partial_d\lambda^a)\mathcal R_c.
 \label{eq:Ba_second_derivative_dictionary}
\end{align}
Terms proportional to $\mathcal B$ vanish only after differentiation. Any smooth coefficient is evaluated on the displaced sheet as
\begin{equation}
 F(r,\theta_\Sigma)=F_0(r)+\chi C\cos\Phi\left.\partial_\theta F\right|_{\pi/2}
 +O(\chi^2,a_\star^4\chi),\qquad F_0=F|_{\pi/2}.
 \label{eq:coefficient_sheet_expansion}
\end{equation}

\subsection{Surface-velocity terms before projection}

We now apply the dictionary to the four terms in
Eq.~\eqref{eq:full_surface_velocity}. The polar and azimuthal advective pieces are immediately
\begin{align}
 \mathcal U_\theta(\Phi)
 &\equiv V^\theta_\Sigma,
 \label{eq:Utheta_preprojection}
 \\
 \mathcal U_\phi(\Phi)
 &\equiv
 (V^\phi-\OmF)\mathcal R_\phi
 =\chi C(V^\phi-\OmF)\sin\Phi+O(\chi^2,a_\star^4\chi).
 \label{eq:Uphi_preprojection}
\end{align}
For the radial-flow term, use $B^a=\lambda^a\mathcal B$ before setting
$\mathcal B=0$.  The derivatives of
$\sqrt\gamma V^r\lambda^a$ multiply $\mathcal B$ and therefore vanish
on the reversal, leaving
\begin{equation}
 {
 \mathcal U_r(\Phi)
 \equiv
 -\frac{\partial_a(\sqrt\gamma V^rB^a)}
 {\sqrt\gamma\,\partial_\theta\mathcal B}
 =-V^r\left(\lambda^\theta
 +\lambda^\phi\mathcal R_\phi\right).}
 \label{eq:Ur_preprojection_exact}
\end{equation}
Using Eqs.~\eqref{eq:lambda_expansion} and
\eqref{eq:phi_theta_ratio} on
$\theta_\Sigma-\pi/2=\chi C\cos\Phi$ gives
\begin{equation}
 \mathcal U_r(\Phi)
 =-\chi C V^r
 \left[\ell\cos\Phi+\lambda_0^\phi\sin\Phi\right]
 +O(\chi^2). \label{eq:Ur_preprojection_harmonics}
\end{equation}
For the magnetic-diffusion term it is useful to factor every magnetic
component by the same sheet marker.  Define
\begin{equation}
 \Lambda^i\equiv(1,\lambda^\theta,\lambda^\phi),
 \qquad
 \mathcal G^i{}_j
 \equiv
 \partial_j(\Nlapse\Lambda^i)
 +\Nlapse\,{}^{(3)}\Gamma^i{}_{jk}\Lambda^k,
 \label{eq:Lambda_G_definition}
\end{equation}
so that
\begin{equation}
 D_j(\Nlapse B^i)
 =\Nlapse\Lambda^i\partial_j\mathcal B
 +\mathcal G^i{}_j\mathcal B.
 \label{eq:covariant_B_derivative_factorized}
\end{equation}
The complete diffusion flux in Eq.~\eqref{eq:Cdiff_general_definition} can
then be written
\begin{equation}
 D^a(\Nlapse\mathcal B)-D^r(\Nlapse B^a)
 =\Psi^a{}_j\partial_j\mathcal B+\Upsilon^a\mathcal B,
 \label{eq:diffusion_flux_factorized}
\end{equation}
where
\begin{equation}
 {
 \Psi^a{}_j
 =\Nlapse\left(\gamma^{aj}-\lambda^a\gamma^{rj}\right),
 \qquad
 \Upsilon^a
 =\gamma^{aj}\mathcal G^r{}_j
 -\gamma^{rj}\mathcal G^a{}_j.}
 \label{eq:Xi_Upsilon_definition}
\end{equation}
The term proportional to $\mathcal B$ vanishes on the sheet only after the
outer derivative has acted.  Dividing by $\partial_\theta\mathcal B$ therefore
gives the full pre-projection diffusion velocity
\begin{equation}
 {
 \mathcal U_{\rm diff}(\Phi)
 \equiv-\frac{\mathcal C_{\rm diff}}{\partial_\theta\mathcal B}
 =-\frac1{\sqrt\gamma}\sum_{a=\theta,\phi}\left[
 \partial_a(\sqrt\gamma\eta_{\rm m}\Psi^a{}_j)\mathcal R_j
 +\sqrt\gamma\eta_{\rm m}\Psi^a{}_j\mathcal R_{aj}
 +\sqrt\gamma\eta_{\rm m}\Upsilon^a\mathcal R_a
 \right]_\Sigma .}
 \label{eq:Udiff_preprojection_dictionary}
\end{equation}
Repeated $j\in\{r,\theta,\phi\}$ is summed.  This equation contains the
entire magnetic-diffusion operator; no angular/vector decomposition has been
made. In the spherical coordinate representation of the Kerr geometry used below,
the inverse spatial metric satisfies
\begin{equation}
 \gamma^{r\theta}=\gamma^{\theta\phi}=0,
\end{equation}
while $\gamma^{r\phi}$ is in general nonzero.  With this coordinate
specialization, the coefficients become
\begin{equation}
 \begin{aligned}
 \Psi^\theta{}_r&=-\Nlapse\lambda^\theta\gamma^{rr},
 &\Psi^\theta{}_\theta&=\Nlapse\gamma^{\theta\theta},
 &\Psi^\theta{}_\phi&=-\Nlapse\lambda^\theta\gamma^{r\phi},\\
 \Psi^\phi{}_r&=\Nlapse(\gamma^{r\phi}-\lambda^\phi\gamma^{rr}),
 &\Psi^\phi{}_\theta&=0,
 &\Psi^\phi{}_\phi&=\Nlapse(\gamma^{\phi\phi}-\lambda^\phi\gamma^{r\phi}),
 \end{aligned}
 \label{eq:Kerr_Xi_coefficients}
\end{equation}
and
\begin{equation}
 {
 \begin{aligned}
 \Upsilon^\theta={}&
 \gamma^{\theta\theta}\mathcal G^r{}_\theta
 -\gamma^{rr}\mathcal G^\theta{}_r
 -\gamma^{r\phi}\mathcal G^\theta{}_\phi,
\\
 \Upsilon^\phi={}&
 \gamma^{r\phi}\mathcal G^r{}_r
 +\gamma^{\phi\phi}\mathcal G^r{}_\phi
 -\gamma^{rr}\mathcal G^\phi{}_r
 -\gamma^{r\phi}\mathcal G^\phi{}_\phi .
 \end{aligned}}
 \label{eq:Kerr_Upsilon_coefficients}
\end{equation}
Axisymmetry removes the $\partial_\phi$ derivatives of these smooth coefficients, but not the $\phi$ derivatives of the tilted sheet marker. 
Equation~\eqref{eq:Udiff_preprojection_dictionary}, together with the local
derivative ratios, is the starting point for the later Kerr expansion.

\subsection{The $m=1$ bookkeeping and the plasma-dependent remainder}
Let us define cosine and sine moments by
\begin{equation}
 [f]_{mc}=\frac1\pi\int_0^{2\pi}f\cos(m\Phi)\,\dd\Phi,
 \qquad
 [f]_{ms}=\frac1\pi\int_0^{2\pi}f\sin(m\Phi)\,\dd\Phi,
 \qquad
 f_0=\frac1{2\pi}\int_0^{2\pi}f\,\dd\Phi .
 \label{eq:Fourier_moments_general}
\end{equation}
The equation for the inclination angle $\chi$ then follows from taking the $m=1$ moment of \eqref{eq:full_surface_velocity}, and its right-hand side is given by two distinct contributions
\begin{equation}
 {
 \frac{\dd\chi}{\dd t}
 =\mathfrak A_{\rm pl}+\mathfrak R_{\rm diff},}
 \label{eq:m1_ledger}
\end{equation}
where $\mathfrak R_{\rm diff}=C^{-1}[\mathcal U_{\rm diff}]_{1c}$ because $\DtPhi\theta_\Sigma=C\dot\chi\cos\Phi$ to linear order in $\chi$.

\subsection{The slow-spin Blandford-Znajek monopole: from Boyer--Lindquist to Kerr--Schild}
\label{sec:BZprofile}
Let us now specify the spacetime and exterior magnetosphere used in the calculation. We do not rederive the Blandford--Znajek solution here, but instead take as given the stationary slow-spin monopole solution obtained from the force-free boundary-value problem. Our purpose is to express this known solution, originally written in Boyer--Lindquist coordinates, in horizon-regular ingoing Kerr--Schild coordinates, where its regularity at the future horizon is manifest. We define
\begin{equation}
 \Delta=r^2-2Mr+a^2,
 \qquad
 \Sigma=r^2+a^2\cos^2\theta,
 \qquad
 r_H=M+\sqrt{M^2-a^2},
 \qquad
 \Omega_H=\frac{a}{r_H^2+a^2}.
 \label{eq:Kerr_BL_definitions}
\end{equation}
In Boyer--Lindquist coordinates
$x^\mu_{\rm BL}=(t_{\rm BL},r,\theta,\phi_{\rm BL})$, the Kerr metric is
\begin{equation}
 {
 \begin{aligned}
 \dd s^2={}&
 -\left(1-\frac{2Mr}{\Sigma}\right)\dd t_{\rm BL}^2
 -\frac{4Mar\sin^2\theta}{\Sigma}\,
   \dd t_{\rm BL}\dd\phi_{\rm BL}
 +\frac{\Sigma}{\Delta}\dd r^2
 +\Sigma\,\dd\theta^2
 \\
 &+
 \frac{\left[(r^2+a^2)^2-a^2\Delta\sin^2\theta\right]
       \sin^2\theta}{\Sigma}\,
 \dd\phi_{\rm BL}^2 .
 \end{aligned}}
 \label{eq:Kerr_BL_metric}
\end{equation}
The factor $\Sigma/\Delta$ makes the Boyer--Lindquist coordinates
singular at $\Delta=0$, although the spacetime itself is regular at the
future horizon. The Blandford-Znajek (BZ) electromagnetic field is reconstructed from the flux function $\psi$, polar current $I$, and field-line angular velocity $\OmF$ as \cite{Blandford1977,McKinney_2004,tanabe,Armas_2020,GrallaJacobson2014}
\begin{equation}
 {
 F_{\rm BL}
 =\frac{I\Sigma}{2\pi\Delta\sin\theta}\,\dd r\wedge\dd\theta
 +\dd\psi\wedge(\dd\phi_{\rm BL}-\OmF\dd t_{\rm BL}).}
 \label{eq:Armas_BL_field}
\end{equation}
For $a_\star=a/M$ and $x=r/M$, the solution needed at
$O(a_\star^2)$ is
\begin{equation}
 {
 \psi
 =\psi_0(1-\cos\theta)
 +a_\star^2\psi_0\Rhat(x)\sin^2\theta\cos\theta
 +O(a_\star^4),}
 \label{eq:BZ_flux}
\end{equation}
while only the leading current and angular velocity are needed,
\begin{equation}
 {
 \OmF=\frac{a_\star}{8M}+O(a_\star^3/M),
 \qquad
 I=-2\pi\OmF\psi_0\sin^2\theta
 +O(a_\star^3\psi_0/M).}
 \label{eq:BZ_I_Omega}
\end{equation}
The radial function is
\begin{align}
 \Rhat(x)=\frac{1}{72x}\Bigg[&
 24+11x+36x^2-36x^3
 +(6x+18x^2-36x^3)\ln\frac{x}{2}
 \nonumber\\
 &+(27x^3-18x^4)\ln\frac{x}{2}\ln\frac{x-2}{x}
 +9x^3(-3+2x)\operatorname{Li}_2\!\left(\frac2x\right)
 \Bigg].
 \label{eq:Rhat_explicit}
\end{align}
The ingoing Kerr--Schild coordinates used below obey
\begin{equation}
 \dd t_{\rm BL}=\dd t-\frac{2Mr}{\Delta}\dd r,
 \qquad
 \dd\phi_{\rm BL}=\dd\phi-\frac{a}{\Delta}\dd r.
 \label{eq:BL_to_KS_exact}
\end{equation}
Applying this transformation to the Boyer--Lindquist metric gives the
Kerr metric in spherical ingoing Kerr--Schild coordinates
$x^\mu=(t,r,\theta,\phi)$:
\begin{equation}
 {
 \begin{aligned}
 \dd s^2={}&
 -\left(1-\frac{2Mr}{\Sigma}\right)\dd t^2
 +\frac{4Mr}{\Sigma}\,\dd t\,\dd r
 -\frac{4Mar\sin^2\theta}{\Sigma}\,\dd t\,\dd\phi
 \\
 &+\left(1+\frac{2Mr}{\Sigma}\right)\dd r^2
 -2a\left(1+\frac{2Mr}{\Sigma}\right)
   \sin^2\theta\,\dd r\,\dd\phi
 +\Sigma\,\dd\theta^2
 \\
 &+\left(
 r^2+a^2
 +\frac{2Ma^2r\sin^2\theta}{\Sigma}
 \right)\sin^2\theta\,\dd\phi^2 .
 \end{aligned}}
 \label{eq:Kerr_KS_metric}
\end{equation}
In contrast to Boyer--Lindquist coordinates, the ingoing Kerr--Schild
metric has nonzero $g_{tr}$ and $g_{r\phi}$ components and is regular
at the future horizon. Substitution of the same coordinate transformation into
Eq.~\eqref{eq:Armas_BL_field} gives
\begin{equation}
 F
 =\frac{\mathcal Z(r,\theta)}{\Delta}\,\dd r\wedge\dd\theta
 +\dd\psi\wedge(\dd\phi-\OmF\dd t),
 \label{eq:KS_field_before_Znajek}
\end{equation}
with
\begin{equation}
 {
 \mathcal Z
 =\frac{I\Sigma}{2\pi\sin\theta}
 -(2Mr\OmF-a)\partial_\theta\psi.}
 \label{eq:Znajek_numerator}
\end{equation}
The apparent $1/\Delta$ pole is removed by the Znajek
condition
\begin{equation}
 I_H
 =2\pi(\OmF-\Omega_H)
 \frac{r_H^2+a^2}{\Sigma_H}
 \sin\theta\,\partial_\theta\psi_H.
 \label{eq:Znajek_condition}
\end{equation}
Indeed, using $2Mr_H=r_H^2+a^2$ and
$a=(r_H^2+a^2)\Omega_H$,
\begin{align}
 \mathcal Z_H
 &=(r_H^2+a^2)(\OmF-\Omega_H)\partial_\theta\psi_H
 -(2Mr_H\OmF-a)\partial_\theta\psi_H
 =0.
 \label{eq:Znajek_explicit_cancellation}
\end{align}
Thus the Boyer--Lindquist singularity is only a coordinate singularity, and
the transformed field is regular on the future horizon. Using Eq.~\eqref{eq:BZ_I_Omega} in
Eq.~\eqref{eq:Znajek_numerator}, and retaining the magnetic field to the order required here,
\begin{align}
 \mathcal Z
 &=\psi_0\sin\theta\,[a-\OmF r(r+2M)]+O(a_\star^3)
 =-\OmF(r-2M)(r+4M)\psi_0\sin\theta+O(a_\star^3).
 \label{eq:Znajek_numerator_slowspin}
\end{align}
Since $\Delta=r(r-2M)+O(a^2)$, this gives
\begin{equation}
 \frac{\mathcal Z}{\Delta}
 =-P(r)\partial_\theta\psi+O(a_\star^3),
 \qquad
 {P(r)=\OmF\left(1+\frac{4M}{r}\right).}
 \label{eq:regular_phase_gradient}
\end{equation}
Consequently, to the order needed for the present magnetic operator,
\begin{equation}
 {
 F=\dd\psi\wedge\dd\Phi+O(a_\star^3),
 \qquad
 \dd\Phi=\dd\phi-\OmF\dd t+P(r)\dd r,}
 \label{eq:Euler_potential_KS}
\end{equation}
with the integrated phase
\begin{equation}
 \Phi=\phi-\OmF\left[t-r-4M\ln\left(\frac{r}{2M}\right)\right]-\Phi_0.
 \label{eq:Schwarzschild_phase_KS}
\end{equation}
which corresponds to the phase that was introduced earlier in \eqref{eq:general_outgoing_phase} as the one characterizing corotation of the current sheet, with 
\begin{align}
    \mathcal M(r)   =   r  + 4M\ln\left(\frac{r}{2M}\right)   ~~ . 
\end{align}

\subsection{Ingoing Kerr--Schild $3+1$ geometry}

Writing Eq.~\eqref{eq:Kerr_KS_metric} in the standard $3+1$ form
\begin{equation}
 \dd s^2
 =-\Nlapse^2\dd t^2
 +\gamma_{ij}(\dd x^i+\beta^i\dd t)
                 (\dd x^j+\beta^j\dd t),
\end{equation}
one obtains
\begin{equation}
 \Nlapse=\left(1+\frac{2Mr}{\Sigma}\right)^{-1/2},
 \qquad
 \sqrt\gamma=\frac{\Sigma\sin\theta}{\Nlapse},
 \label{eq:KS_lapse_det}
\end{equation}
and
\begin{equation}
 {
 \beta^i=\left(\frac{2Mr}{\Sigma+2Mr},0,0\right),
 \qquad
 \beta_i=\left(\frac{2Mr}{\Sigma},0,
 -\frac{2Mar\sin^2\theta}{\Sigma}\right).}
 \label{eq:Kerr_shift}
\end{equation}
The nonzero spatial metric components needed here are
\begin{align}
 \gamma_{rr}&=1+\frac{2Mr}{\Sigma},
 &\gamma_{r\phi}&=-a\left(1+\frac{2Mr}{\Sigma}\right)\sin^2\theta,
 \\
 \gamma_{\theta\theta}&=\Sigma,
 &\gamma_{\phi\phi}&=\left(r^2+a^2
 +\frac{2Ma^2r\sin^2\theta}{\Sigma}\right)\sin^2\theta,
\end{align}
and useful inverse components are
\begin{equation}
 \gamma^{r\phi}=\frac{a}{\Sigma},
 \qquad
 \gamma^{\theta\theta}=\frac1\Sigma,
 \qquad
 \gamma^{\phi\phi}=\frac1{\Sigma\sin^2\theta}   ,
 \qquad
 \gamma^{rr}
 =\frac{\Sigma(r^2+a^2)+2Ma^2r\sin^2\theta}
 {\Sigma(\Sigma+2Mr)}.
 \label{eq:Kerr_inverse_rr}
\end{equation}

\subsection{Magnetic components, exterior ratios, and the oblique sheet}

The flux function $\psi$ introduced above describes the smooth, unsplit
monopolar exterior solution; the split-monopole reversal will be imposed
separately through the common profile $\mathcal Q(\Xi)$.  Equation
\eqref{eq:Euler_potential_KS} therefore gives the exterior magnetic field on
a spatial slice as
\begin{equation}
 {
 B_{\rm ext}^r=\frac1{\sqrt\gamma}\partial_\theta\psi,
 \qquad
 B_{\rm ext}^\theta=-\frac1{\sqrt\gamma}\partial_r\psi,
 \qquad
 B_{\rm ext}^\phi=-P(r)B_{\rm ext}^r+O(a_\star^3).}
 \label{eq:B_components_Euler_phase}
\end{equation}
For the common smooth reversal we take
\begin{equation}
 B^i=\mathcal Q(\Xi)B_{\rm ext}^i,
\end{equation}
and the ideal electric field carries the same reversal factor. From $E_{{\rm ext},i}^{\rm id}=F_{i\mu}n^\mu$ and
Eq.~\eqref{eq:Euler_potential_KS},
\begin{equation}
 {
 \begin{aligned}
 \Nlapse E_{{\rm ext},r}^{\rm id}&=\OmF\sqrt\gamma B_{\rm ext}^\theta,
 \\
 \Nlapse E_{{\rm ext},\theta}^{\rm id}
 &=-\big(\OmF+\beta^rP\big)\sqrt\gamma B_{\rm ext}^r,
 \\
 \Nlapse E_{{\rm ext},\phi}^{\rm id}
 &=-\beta^r\sqrt\gamma B_{\rm ext}^\theta.
 \end{aligned}}
 \label{eq:Eid_components_BZ}
\end{equation}
The previously general exterior ratios are now fixed:
\begin{equation} 
 {
 \lambda^\theta=-\frac{\partial_r\psi}{\partial_\theta\psi},
 \qquad
 \lambda^\phi=-P(r).}
 \label{eq:lambda_BZ}
\end{equation} 
The equatorial expansion used earlier in
Eq.~\eqref{eq:lambda_expansion} follows directly. From Eq.~\eqref{eq:BZ_flux},
\begin{align}
 \partial_r\psi
 &=
 \frac{a_\star^2\psi_0}{M}\Rhat'(x)
 \sin^2\theta\cos\theta+O(a_\star^4),
 \\
 \partial_\theta\psi
 &=\psi_0\sin\theta+O(a_\star^2).
 \label{eq:psi_derivatives_for_lambda}
\end{align}
Evaluating on $\theta_\Sigma=\pi/2+\chi C\cos\Phi+\cdots$ gives
\begin{equation}
 \left.\lambda^\theta\right|_\Sigma
 =\frac{a_\star^2}{M}\Rhat'(x)\chi\cos\Phi
 +O(a_\star^2\chi^3/M,a_\star^4\chi/M).
 \label{eq:lambda_theta_equatorial_derived}
\end{equation}

Therefore
\begin{equation}
 {
 \ell(r)=\left.\partial_\theta\lambda^\theta\right|_{\pi/2}
 =\frac{a_\star^2}{M}\Rhat'(x)+O(a_\star^4).}
 \label{eq:ell_BZ}
\end{equation}
Similarly, $B^\phi/B^r=-P$ gives $\lambda^\phi=-P$ pointwise.  The
same Blandford-Znajek data also determine the single exterior amplitude used in the sheet
ansatz.  Since the tracked marker is $\mathcal B=B^r$, its unsigned exterior
amplitude is
\begin{equation}
 {
 A_B(r,\theta)
 =\frac{\Nlapse}{\Sigma\sin\theta}\,\partial_\theta\psi .}
 \label{eq:AB_full_definition}
\end{equation}
With $\Nlapse_0=\sqrt{x/(x+2)}$, Eqs.~\eqref{eq:BZ_flux} and
\eqref{eq:KS_lapse_det} give the complete slow-spin amplitude
\begin{equation}
 A_B
 =\frac{\Nlapse_0\psi_0}{r^2}
 \left[1+a_\star^2\left(
 \Rhat(x)(3\cos^2\theta-1)
 -\frac{x+1}{x^2(x+2)}\cos^2\theta\right)\right]
 +O(a_\star^4),  \label{eq:AB_full_slowspin}
\end{equation}
from which it follows that,
\begin{equation}
 {
 \left.\partial_\theta^2\ln A_B\right|_{\pi/2}
 =a_\star^2\left[
 6\Rhat(x)-\frac{2(x+1)}{x^2(x+2)}
 \right]+O(a_\star^4).}
 \label{eq:AB_full_equatorial_curvature}
\end{equation}
These relations, together with $\ell=a_\star^2\Rhat'/M$, $\lambda^\phi=-P$, and the derivatives of the sheet label, determine the diffusive term below. The magnetic colatitude obeys
\begin{equation}
 \vartheta=\theta+a_\star^2\Rhat(x)\sin\theta\cos\theta+O(a_\star^4).
 \label{eq:magnetic_colatitude}
\end{equation}
It is necessary to differentiate this full angular expression before restricting to the inclined sheet: replacing it by its linear Taylor expansion about the equator would lose the curvature in Eq.~\eqref{eq:Xi_polar_curvature}. The common-profile split uses
\begin{equation}
 \Xi=\cos\chi\cos\vartheta+\sin\chi\sin\vartheta\cos\Phi,
 \qquad B^i=\mathcal Q(\Xi)B^i_{\rm ext}.
 \label{eq:complete_initial_split}
\end{equation}
Its zero has the warped angular location given in Eq.~\eqref{eq:small_alpha_sheet}.

\subsection{Plasma-dependent advection and the ideal-background check}

The velocity-dependent terms form the first sector of
Eq.~\eqref{eq:m1_ledger}.  Force-free electrodynamics determines the
electromagnetic field but not the full plasma transport, so
$\mathfrak A_{\rm pl}$ is kept as an analytically unresolved contribution. It is nevertheless useful to check what these terms do for the stationary ideal Blandford-Znajek configuration about which the resistive calculation is performed.  Before imposing ideality, the three advective terms derived in
Eqs.~\eqref{eq:Utheta_preprojection},
\eqref{eq:Uphi_preprojection}, and
\eqref{eq:Ur_preprojection_exact} combine on the sheet to
\begin{equation}
 {
 \begin{aligned}
 \mathcal U_{\rm adv}
 &\equiv
 \mathcal U_\theta+\mathcal U_\phi+\mathcal U_r =
 V^\theta
 +(V^\phi-\OmF)\mathcal R_\phi
 -V^r\left(\lambda^\theta+\lambda^\phi\mathcal R_\phi\right)
 \\
 &=
 \left(V^\theta-V^r\lambda^\theta\right)
 +\left(V^\phi-\OmF-V^r\lambda^\phi\right)\mathcal R_\phi .
 \end{aligned}}
 \label{eq:advective_preprojection_combined}
\end{equation}
No condition on the velocity has entered this expression. For the stationary Blandford-Znajek background the electromagnetic field has the
Euler-potential form
\begin{equation}
 F=\dd\psi\wedge\dd\Phi .
\end{equation}
If the zeroth-order plasma transport is ideal, then
$\iota_uF=0$.  Contracting the Euler-potential form explicitly gives
\begin{equation}
 \iota_uF
 =
 \iota_u(\dd\psi\wedge\dd\Phi)
 =
 (u^\mu\partial_\mu\psi)\,\dd\Phi
 -(u^\mu\partial_\mu\Phi)\,\dd\psi .
 \label{eq:iuf_Euler_potential_expanded}
\end{equation}
Away from points where $\dd\psi$ and $\dd\Phi$ become linearly
dependent, these are independent one-forms.  Therefore
$\iota_uF=0$ requires the two coefficients separately to vanish,
\begin{equation}
 {
 u^\mu\partial_\mu\psi=0,
 \qquad
 u^\mu\partial_\mu\Phi=0.}
 \label{eq:Euler_potential_transport_constraints}
\end{equation}
Thus an ideal worldline preserves both Euler-potential labels. Using
\begin{equation}
 u^\mu=u^t(1,V^r,V^\theta,V^\phi),
 \qquad
 V^i=\Nlapse v^i-\beta^i ,
 \label{eq:u_coordinate_velocity}
\end{equation}
together with $\psi=\psi(r,\theta)$ and
$\dd\Phi=\dd\phi-\OmF\dd t+P(r)\dd r$, the two conditions in
Eq.~\eqref{eq:Euler_potential_transport_constraints} become
\begin{equation}
 V^r\partial_r\psi+V^\theta\partial_\theta\psi=0,
 \qquad
 V^\phi-\OmF+PV^r=0.
 \label{eq:kernel_conditions_components}
\end{equation}
The magnetic-field ratios derived from the same Blandford-Znajek two-form,
Eq.~\eqref{eq:lambda_BZ}, are
\begin{equation}
 \lambda^\theta
 =-\frac{\partial_r\psi}{\partial_\theta\psi},
 \qquad
 \lambda^\phi=-P .
\end{equation}
Hence Eq.~\eqref{eq:kernel_conditions_components} is equivalently
\begin{equation}
 {
 V^\theta=V^r\lambda^\theta,
 \qquad
 V^\phi-\OmF=V^r\lambda^\phi .}
 \label{eq:ideal_transport_relations_pointwise}
\end{equation}
Substituting only these two ideal-background relations,
Eq.~\eqref{eq:ideal_transport_relations_pointwise}, into the
brute-force pre-projection result
Eq.~\eqref{eq:advective_preprojection_combined} gives
\begin{equation}
 {
 \mathcal U_{\rm adv}
 =
 \underbrace{\left(V^\theta-V^r\lambda^\theta\right)}_{0}
 +
 \underbrace{\left(V^\phi-\OmF-V^r\lambda^\phi\right)}_{0}
 \mathcal R_\phi
 =0.}
 \label{eq:ideal_preprojection_cancellation}
\end{equation}
This result is consistent with relativistic flux freezing,
$\dd F=0$ and $\iota_uF=0\Rightarrow\mathcal L_uF=0$ \cite{Cowling1953SolarElectrodynamics,GrallaJacobson2014}, but we can use it only at ideal order, whereas the actual current sheet is resistive, so its plasma velocity need not continue to obey
Eq.~\eqref{eq:ideal_transport_relations_pointwise}.  Any resulting
velocity-dependent contribution to the subsequent evolution of
$\chi$ belongs to $\mathfrak A_{\rm pl}$ and is outside the
controlled resistive mechanism isolated here.

\subsection{Full magnetic-diffusion sector before projection}
We now evaluate $\mathcal U_{\rm diff}$ from Eq.~\eqref{eq:Udiff_preprojection_dictionary}. Define
\begin{equation}
 L_r=\left.\partial_r\ln A_B\right|_{\pi/2}.
 \label{eq:Lr_definition}
\end{equation}
For the symmetric profile $\mathcal Q''(0)=0$, the required ratios are
\begin{align}
 \mathcal R_r
 &=\chi CP\sin\Phi-\chi C'\cos\Phi+\cdots,\nonumber\\
 \mathcal R_{\theta r}
 &=L_r-C'+O(\chi^2/M,a_\star^4/M),\nonumber\\
 \mathcal R_\phi
 &=\chi C\sin\Phi+\cdots,
 \qquad \mathcal R_{\theta\phi}=O(\chi^2),\nonumber\\
 \mathcal R_{\phi\phi}
 &=\chi C\cos\Phi+\cdots,\nonumber\\
 \mathcal R_{\phi r}
 &=\chi C[P\cos\Phi+L_r\sin\Phi]+\cdots,\nonumber\\
 \mathcal R_{\theta\theta}
 &=2\left.\partial_\theta\ln A_B\right|_\Sigma
 +4(C-1)\chi\cos\Phi+\cdots.
 \label{eq:R_dictionary_symmetric_collected}
\end{align}
Here the ellipses indicate higher spin or inclination orders, with
\begin{equation}
 P(r)=\frac{a_\star}{8M}\left(1+\frac4x\right)+O(a_\star^3/M),
 \qquad \ell=C'=\frac{a_\star^2}{M}\Rhat'(x),\qquad C=1+a_\star^2\Rhat(x).
 \label{eq:P_ell_explicit_diffusion}
\end{equation}
The correction to $\mathcal R_{\theta r}$ multiplies $\Psi^\theta{}_r=O(a_\star^2\chi)$ and contributes only at fourth spin order. The $-\chi C'\cos\Phi$ part of $\mathcal R_r$ likewise first contributes to the cosine sector at fourth order. By contrast, the extra polar curvature in $\mathcal R_{\theta\theta}$ contributes at second order and must be retained.

It is useful as an algebraic device to separate the two outer derivatives
$a=\theta$ and $a=\phi$ in
Eq.~\eqref{eq:Udiff_preprojection_dictionary}, so that 
\begin{align}
 \mathcal U_{\rm diff} =  \mathcal U_{\rm diff}^{(\theta)} +  \mathcal U_{\rm diff}^{(\phi)}  ~~ .   
 \end{align}
For the $\theta$ sector,
it is furthermore useful to keep track of the origin of each term.  We
write
\begin{equation}
 \mathcal U_{\rm diff}^{(\theta)}
 = 
 \sum_{j=r,\theta,\phi}
\Big[ 
 \mathcal C_j^{(\theta)}
 +\mathcal S_j^{(\theta)}
   \Big]_\Sigma
 -\eta_{\rm m}\Upsilon^\theta   |_\Sigma  ,
 \label{eq:Udiff_theta_ledger_definition}
\end{equation}
where
\begin{align}
 \mathcal C_j^{(\theta)}
 &\equiv
 -\frac1{\sqrt\gamma}
 \partial_\theta
 \left(\sqrt\gamma\eta_{\rm m}\Psi^\theta{}_j\right)
 \mathcal R_j,
 \label{eq:Ctheta_definition}
 \\
 \mathcal S_j^{(\theta)}
 &\equiv
 -\eta_{\rm m}\Psi^\theta{}_j\mathcal R_{\theta j},
 \label{eq:Stheta_definition}  
\end{align}
The $\mathcal C_j^{(\theta)}$ terms arise when the outer
$\theta$ derivative acts on the smooth coefficients, while
$\mathcal S_j^{(\theta)}$ contains the second derivatives of the sheet
marker.  Finally, we collect the connection terms contained in
$\Upsilon^\theta$. The $\phi$ sector is given by
\begin{equation}
 \mathcal U_{\rm diff}^{(\phi)}
 =-\eta_{\rm m}\Big[
 \Psi^\phi{}_r\mathcal R_{\phi r}
 +\Psi^\phi{}_\phi\mathcal R_{\phi\phi}
 +\Upsilon^\phi\mathcal R_\phi
 \Big]_\Sigma .
 \label{eq:Udiff_phi_exact}
\end{equation}
We now evaluate the seven $\theta$-sector terms at the order relevant for the $m=1$ cosine projection. All coefficients are differentiated first and then evaluated at $\theta_\Sigma=\pi/2+\chi C\cos\Phi+\cdots$. From Eq.~\eqref{eq:Kerr_Xi_coefficients},
\begin{equation}
 \left.\Psi^\theta{}_r\right|_\Sigma
 =-\frac{a_\star^2}{M}\alpha_0^3\Rhat'\chi\cos\Phi+\cdots,
 \qquad \Psi^\theta{}_\theta=\frac{\Nlapse}{\Sigma},
 \qquad \left.\Psi^\theta{}_\phi\right|_\Sigma=O(a_\star^3\chi/M^2).
 \label{eq:Xitheta_phi_order}
\end{equation}
This power counting removes three terms from the $O(a_\star^2\chi\cos\Phi)$ coefficient. The sine part of $\mathcal C_r^{(\theta)}$ starts at $O(a_\star^3\chi)$ and its cosine part at $O(a_\star^4\chi)$. The term $\mathcal C_\phi^{(\theta)}$ is also beyond the retained spin order, while $\mathcal S_\phi^{(\theta)}$ is additionally suppressed by $\mathcal R_{\theta\phi}=O(\chi^2)$.

Thus the $\theta$-sector cosine
coefficient through $O(a_\star^2\chi)$ receives contributions only from
\begin{equation}
 \mathcal C_\theta^{(\theta)},\qquad
 \mathcal S_\theta^{(\theta)},\qquad
 \mathcal S_r^{(\theta)},\qquad
 -\eta_{\rm m}\Upsilon^\theta   .
 \label{eq:theta_four_surviving_terms}
\end{equation}
We evaluate these four terms in turn. The first surviving term is
$\mathcal C_\theta^{(\theta)}$.  Using
$\sqrt\gamma=\Sigma\sin\theta/\Nlapse$ and
$\Psi^\theta{}_\theta=\Nlapse/\Sigma$ gives the particularly simple identity
\begin{equation} \sqrt\gamma\,\Psi^\theta{}_\theta=\sin\theta .
\end{equation}
Hence
\begin{align}
 \mathcal C_\theta^{(\theta)}
 &=
 -\frac1{\sqrt\gamma}
 \partial_\theta
 \left(\sqrt\gamma\eta_{\rm m}\Psi^\theta{}_\theta\right) =
 -\frac{\eta_{\rm m}}{\sqrt\gamma}\cos\theta .
\end{align}
Evaluating on the displaced sheet,
$\cos\theta_\Sigma=-\chi C\cos\Phi+O(\chi^3,a_\star^4\chi)$, gives
\begin{equation}
 \left.\mathcal C_\theta^{(\theta)}\right|_{\cos\Phi}
 =
 +\frac{\eta_{\rm m}\alpha_0}{M^2x^2}\chi C\cos\Phi
 +O(\chi^3,a_\star^4\chi).
 \label{eq:theta_weight_first_piece}
\end{equation}
Its zeroth-spin part cancels against the azimuthal sector, as required by rotational symmetry. The identical factor $C(r)$ must be retained in both terms, so its second-order correction cancels as well. The term $\mathcal S_\theta^{(\theta)}$ contains the polar curvature of the reversing field. For $\mathcal Q''(0)=0$, Eq.~\eqref{eq:theta_derivative_ratio} gives
\begin{equation}
 \mathcal R_{\theta\theta}
 =2\left.\partial_\theta\ln A_B\right|_\Sigma
 +\left.\frac{\Xi_{\theta\theta}}{\Xi_\theta}\right|_\Sigma.
\end{equation}
Using reflection symmetry of $A_B$ and Eq.~\eqref{eq:Xi_polar_curvature},
\begin{equation}
 \mathcal R_{\theta\theta}
 =2\chi\cos\Phi\left.\partial_\theta^2\ln A_B\right|_{\pi/2}
 +4a_\star^2\Rhat\chi\cos\Phi+O(a_\star^4\chi,\chi^3).
 \label{eq:Rthetatheta_from_amplitude}
\end{equation}

Using Eq.~\eqref{eq:AB_full_equatorial_curvature} and the leading value
$\Psi^\theta{}_{\theta,0}=\alpha_0/(M^2x^2)$ therefore yields
\begin{equation}
 \left.\mathcal S_\theta^{(\theta)}\right|_{\cos\Phi}
 =
 -\frac{\eta_{\rm m}\alpha_0}{M^2x^2}
 \mathcal R_{\theta\theta}
 =
 -\frac{\eta_{\rm m}a_\star^2}{M^2}\chi\cos\Phi
 \left[
 \frac{16\alpha_0}{x^2}\Rhat
 -\frac{4 \alpha_0(x+1)}{x^4(x+2)}
 \right].
 \label{eq:full_amplitude_piece_inside_diffusion}
\end{equation}
Let us now consider the term $-\eta_{\rm m}\Upsilon^\theta$. Equatorial reflection symmetry makes the polar component odd, so
\begin{equation}
 \left.\Upsilon^\theta\right|_0=0.
 \label{eq:Upsilontheta_equatorial_zero}
\end{equation}
Using the displaced-sheet expansion
Eq.~\eqref{eq:coefficient_sheet_expansion},
\begin{align}
 \left.\Upsilon^\theta\right|_\Sigma
 &=
 (\theta_\Sigma-\pi/2)
 \left.\partial_\theta\Upsilon^\theta\right|_0
 +O(\chi^2) =
 \chi\cos\Phi\,
 \left.\partial_\theta\Upsilon^\theta\right|_0
 +O(a_\star^4\chi,\chi^2),
 \label{eq:Upsilontheta_displaced_sheet}
\end{align}
where the $O(a_\star^2\chi)$ correction to $\theta_\Sigma-\pi/2$ can be dropped in
this term because the slope computed below is itself $O(a_\star^2)$. From
Eqs.~\eqref{eq:Lambda_G_definition} and
\eqref{eq:Kerr_Upsilon_coefficients},
\begin{equation}
 \left.\partial_\theta\Upsilon^\theta\right|_0
 =
 \left[
 \partial_\theta\!\left(\gamma^{\theta\theta}\mathcal G^r{}_\theta\right)
 -\partial_\theta\!\left(\gamma^{rr}\mathcal G^\theta{}_r\right)
 -\partial_\theta\!\left(\gamma^{r\phi}\mathcal G^\theta{}_\phi\right)
 \right]_0 .
 \label{eq:dUpsilontheta_split}
\end{equation}
Thus this single term consists of three geometric pieces.  Using the
Kerr--Schild metric coefficients together with
$\lambda^\theta=-(a_\star^2/M)\Rhat'\sin\theta\cos\theta+O(a_\star^4/M)$ and
$\lambda^\phi=-P$, they are
\begin{align}
 \left.
 \partial_\theta\!\left(\gamma^{\theta\theta}\mathcal G^r{}_\theta\right)
 \right|_0
 &=
 \frac{a_\star^2}{M^2} \alpha_0^3
 \left[
 -\frac{\Rhat'}{x}
 +\frac{(x+2)^2}{x^6}
 \right],
 \nonumber\\
 -\left.
 \partial_\theta\!\left(\gamma^{rr}\mathcal G^\theta{}_r\right)
 \right|_0
 &=
 \frac{a_\star^2}{M^2} \alpha_0^3
 \left[
 -\Rhat''
 -\frac{x+3}{x(x+2)}\Rhat'
 -\frac{x^3+6x^2+8x+16}{8x^5}
 \right],
 \nonumber\\
 -\left.
 \partial_\theta\!\left(\gamma^{r\phi}\mathcal G^\theta{}_\phi\right)
 \right|_0
 &=
 \frac{a_\star^2}{M^2} \alpha_0^3
 \frac{(x+2)(x^3+4x^2+8x+16)}{8x^6}.
 \label{eq:dUpsilontheta_three_pieces}
\end{align}
Adding the three pieces gives
\begin{equation}
 {
 \left.\partial_\theta\Upsilon^\theta\right|_0
 =-\frac{a_\star^2}{M^2} \alpha_0^3
 \left[
 \Rhat''+\frac{2x+5}{x(x+2)}\Rhat'
 -\frac{2(x^2+3x+4)}{x^6}
 \right].}
 \label{eq:dUpsilontheta_explicit}
\end{equation}
Therefore
\begin{equation}
  -\eta_{\rm m}\Upsilon^\theta   |_{\cos\Phi}
 =
 \frac{\eta_{\rm m}a_\star^2}{M^2} \alpha_0^3\chi\cos\Phi
 \left[
 \Rhat''+\frac{2x+5}{x(x+2)}\Rhat'
 -\frac{2(x^2+3x+4)}{x^6}
 \right].
 \label{eq:Upsilontheta_cos}
\end{equation}
The final surviving term is
$\mathcal S_r^{(\theta)}
=-\eta_{\rm m}\Psi^\theta{}_r\mathcal R_{\theta r}$. It follows from \eqref{eq:R_dictionary_symmetric_collected} that
\begin{equation}
 \mathcal R_{\theta r}=L_r-\frac{a_\star^2}{M}\Rhat'+O(\chi^2/M,a_\star^4/M).
\end{equation}
The second-order correction in this ratio contributes only at fourth order after multiplication by $\Psi^\theta{}_r$. At leading order,
\begin{equation}
 L_r   =\frac1M\left[
 \frac1{x(x+2)}-\frac2x
 \right]+O(a_\star^2/M),
 \label{eq:Lr_leading_explicit}
\end{equation}
while evaluation of $\Psi^\theta{}_r$ on
$\theta_\Sigma=\pi/2+\chi C\cos\Phi+\cdots$ gives
\begin{equation}
 \left.\Psi^\theta{}_r\right|_\Sigma
 =
 -\frac{a_\star^2}{M} \alpha_0^3\Rhat'\,
 \chi\cos\Phi
 +O(a_\star^4\chi,\chi^2).
\end{equation}
Consequently
\begin{equation}
 \left.
 \mathcal S_r^{(\theta)}
 \right|_{\cos\Phi}
 =
 \left.
 -\eta_{\rm m}\Psi^\theta{}_r\mathcal R_{\theta r}
 \right|_{\cos\Phi}
 =
 -\frac{\eta_{\rm m}a_\star^2}{M^2} \alpha_0^3
 \frac{2x+3}{x(x+2)}\Rhat'\,
 \chi\cos\Phi .
 \label{eq:Xitheta_r_Lr_cos}
\end{equation}
The $\Rhat'$ terms from
$-\eta_{\rm m}\Upsilon^\theta    $ and $\mathcal S_r^{(\theta)}$ combine as
\begin{equation}
 \frac{2x+5}{x(x+2)}
 -\frac{2x+3}{x(x+2)}
 =
 \frac{2}{x(x+2)}.
\end{equation}
Collecting the four surviving contributions
$\mathcal C_\theta^{(\theta)}$,
$\mathcal S_\theta^{(\theta)}$,
$\mathcal S_r^{(\theta)}$, and
$ -\eta_{\rm m}\Upsilon^\theta   $, we thus obtain
\begin{equation}
 \begin{aligned}
 \left.\mathcal U_{\rm diff}^{(\theta)}\right|_{\cos\Phi}
 ={}&
 +\frac{\eta_{\rm m} \alpha_0}{M^2x^2}\chi C\cos\Phi
 -\frac{\eta_{\rm m}a_\star^2}{M^2}\chi\cos\Phi
 \left[
 \frac{16 \alpha_0}{x^2}\Rhat
 -\frac{4 \alpha_0(x+1)}{x^4(x+2)}
 \right]
 \\
 &+\frac{\eta_{\rm m}a_\star^2}{M^2} \alpha_0^3\chi\cos\Phi
 \left[
 \Rhat''+\frac{2\Rhat'}{x(x+2)}
 -\frac{2(x^2+3x+4)}{x^6}
 \right].
 \end{aligned}
 \label{eq:Udiff_theta_cos_explicit}
\end{equation}
Let us now consider $  \mathcal U_{\rm diff}^{(\phi ) }$. For this, we use that from \eqref{eq:Kerr_Xi_coefficients} follow the relations
\begin{align}
 \left.\Psi^\phi{}_r\right|_0
 &=\frac{a_\star}{M} \alpha_0
 \left[
 \frac{ \alpha_0^2}{8}\left(1+\frac4x\right)+\frac1{x^2}
 \right]+O(a_\star^3/M),
 \label{eq:Xiphi_r_explicit}\\
 \left.\Psi^\phi{}_\phi\right|_0
 &=\frac{ \alpha_0}{M^2x^2}
 +\frac{a_\star^2}{M^2}\frac{ \alpha_0}{8x^2}
 \left(1+\frac4x\right)
 +O(a_\star^4/M^2).
 \label{eq:Xiphi_phi_explicit}
\end{align} 
In the expression given in \eqref{eq:Udiff_phi_exact}, there is also a term $- \eta_{\rm m} \Upsilon^\phi \mathcal{R}_\phi $. It follows from \eqref{eq:R_dictionary_symmetric_collected} that $\mathcal R_\phi=\chi C\sin\Phi+\cdots$, which means that this term does not contribute to the moment $m=1 $ at $\mathcal{O} ( a_\star^2  \chi )$, as the term $\Upsilon^\phi $ cannot undo the sine at zeroth order in $\chi $. Substituting the expressions for $\mathcal R_{\phi r}$ and $\mathcal  R_{\phi\phi}
$ then turns \eqref{eq:Udiff_phi_exact} into
\begin{equation}
 {
 \begin{aligned}
 \left.\mathcal U_{\rm diff}^{(\phi)}\right|_{\cos\Phi}
 ={}&-\frac{\eta_{\rm m} \alpha_0}{M^2x^2}\chi C\cos\Phi -\frac{\eta_{\rm m}a_\star^2}{M^2}\chi\cos\Phi
 \left[
 \frac{ \alpha_0^3}{64}\left(1+\frac4x\right)^2
 +\frac{ \alpha_0}{4x^2}\left(1+\frac4x\right)
 \right].
 \end{aligned}}
 \label{eq:Udiff_phi_cos_explicit}
\end{equation}
Adding the remaining $O(a_\star^2)$ terms exposes a second useful
cancellation.  We find
\begin{equation}
  \alpha_0^3\frac{2(x^2+3x+4)}{x^6}
 +\frac{ \alpha_0}{4x^2}\left(1+\frac4x\right)
 = \alpha_0^3\frac{x^4+6x^3+16x^2+24x+32}{4x^6},
\end{equation}
and
\begin{equation}
  \alpha_0^3\frac{x^4+6x^3+16x^2+24x+32}{4x^6}
 -\frac{4 \alpha_0(x+1)}{x^4(x+2)}
 = \alpha_0^3\frac{x^4+6x^3+8x+32}{4x^6}.
 \label{eq:diffusion_geometric_cancellation}
\end{equation}
Combining Eqs.~\eqref{eq:Udiff_theta_cos_explicit} and \eqref{eq:Udiff_phi_cos_explicit}, the terms proportional to $\chi C$ cancel. Taking the $m=1$ cosine moment and dividing by $C$ changes the surviving second-order terms only at fourth order. The resulting rate is 
\begin{equation} 
 \Gamma_{\rm diff}
 =\frac{\eta_{\rm m}a_\star^2}{M^2}\left[
 \frac{16 \alpha_0}{x^2}\Rhat
 +\frac{ \alpha_0^3}{64}\left(1+\frac4x\right)^2
 - \alpha_0^3\left(\Rhat''+\frac{2\Rhat'}{x(x+2)}\right)
 + \alpha_0^3\frac{x^4+6x^3+8x+32}{4x^6}
 \right] , 
 \label{eq:Gamma_diff_complete}
\end{equation}
with 
\begin{equation}
 {
 \mathfrak R_{\rm diff}
 =-\Gamma_{\rm diff}\chi.}
 \label{eq:diffusion_moment_reduced}
\end{equation}
\subsection{Horizon value}
Because the controlled rate is already proportional to $a_\star^2$, the location of the Kerr horizon may itself be expanded at leading order:
\begin{equation}
 \frac{r_H}{M}
 =1+\sqrt{1-a_\star^2}
 =2-\frac{a_\star^2}{2}+O(a_\star^4).
\end{equation}
Writing
\begin{equation}
 \Gamma_{\rm diff}(x)
 =\frac{\eta_{\rm m}a_\star^2}{M^2}\,\mathcal G(x),
\end{equation}
one has
$\mathcal G(x_H)=\mathcal G(2)+O(a_\star^2)$, so the spin correction to
the horizon position affects the rate only at $O(\eta_{\rm m}a_\star^4/M^2)$.
It is therefore consistent to evaluate every coefficient in
Eq.~\eqref{eq:Gamma_diff_complete} at the Schwarzschild horizon $x=2$.
There,
\begin{equation}
 \Rhat(2)=0.14191148,
 \qquad
 \Rhat'(2)=-0.07426557,
 \qquad
 \Rhat''(2)=0.07610165,
 \qquad
  \alpha_0(2)=\frac1{\sqrt2}.
\end{equation}
For the prescribed radial sheet shape in Eq.~\eqref{eq:small_alpha_sheet}, the local horizon rate is
\begin{equation}
 {
 \Gamma_H
 =0.58544\,\frac{\eta_{\rm m}a_\star^2}{M^2}
 +O\!\left(\eta_{\rm m}a_\star^4/M^2\right).}
 \label{eq:Gamma_horizon_numeric}
\end{equation}
Hence the horizon alignment law predicted by the controlled
magnetic-diffusion mechanism is
\begin{equation}  \left.\frac{\dd\chi}{\dd t}\right|_H
  \approx  -0.58544\,\frac{\eta_{\rm m}a_\star^2}{M^2}\,\chi  .
 \label{eq:horizon_alignment_final}
\end{equation}

\section{Inclined current sheet}
\label{app:warped_sheet}
In this section we extend the striped-wind current sheet computation of Bogovalov \cite{Bogovalov:1999vg} to the case of slow-spin black-hole magnetospheres and uncover an additional warping effect for the current sheet. We then verify that this warped current sheet allows for the magnetic Gauss law of an Eulerian observer to be obeyed.
\subsection{Striping and warping}
\begin{figure}
 \centering
\includegraphics[width=0.5\linewidth]
{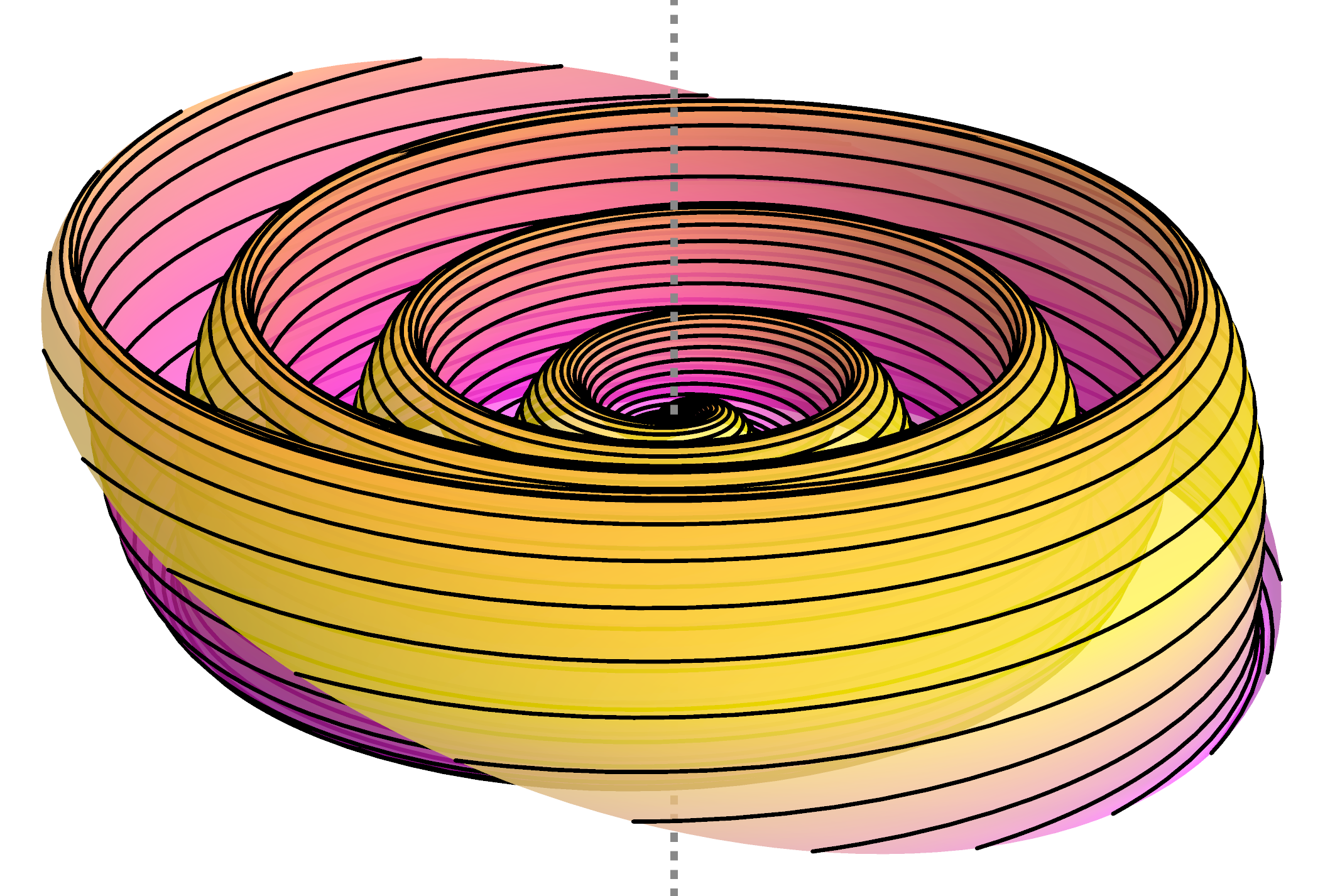}
 \caption{\label{fig:warped_sheet}
 Inclined BZ current sheet used to initialize the
 simulations.  The radial phase delay in
 Eq.~\eqref{eq:main_BZ_phase} winds the sheet into a striped
 structure, while the $O(a_\star^2)$ deformation of the BZ flux
 surfaces generates the additional radial warp in
 Eq.~\eqref{eq:main_warped_sheet}.  The dotted line denotes the
 black-hole spin axis.}
\end{figure}

The preceding Section~\ref{app:derivation} specifies the smooth slow-spin Blandford--Znajek
monopole and shows that, in ingoing Kerr--Schild coordinates and to the
order required here, its electromagnetic field may be written as \cite{GrallaJacobson2014}
\begin{equation}
 F=\dd\psi\wedge\dd\Phi+O(a_\star^3),
 \label{eq:appB_Euler}
\end{equation}
with $\psi$ given by Eq.~\eqref{eq:BZ_flux} and the outgoing phase
$\Phi$ by Eq.~\eqref{eq:Schwarzschild_phase_KS}.  We now use these two
fields to determine the geometry of an oblique split
monopole. Let the smooth Blandford-Znajek field be reversed across a surface $\Xi=0$,
\begin{equation}
 \overline F=\operatorname{sgn}(\Xi)F .
 \label{eq:appB_sign_split}
\end{equation}
It follows that 
\begin{equation}
 \dd\overline F
 =
 2\delta(\Xi)\,\dd\Xi\wedge F
 +\operatorname{sgn}(\Xi)\,\dd F .
\end{equation}
Since the unsplit Blandford-Znajek field obeys $\dd F=0$, the homogeneous Maxwell
equation is preserved provided \cite{GrallaJacobson2014}
\begin{equation}
 \dd\Xi\wedge F=0
 \qquad\text{on }\Xi=0 .
 \label{eq:appB_no_monopole}
\end{equation}
With Eq.~\eqref{eq:appB_Euler}, this condition is satisfied by any
surface of the form
\begin{equation}
 \Xi=\Xi(\psi,\Phi).
 \label{eq:appB_sheet_labels}
\end{equation}
Thus, for fixed $\chi$, the current sheet is transported along the field sheets of the unsplit Blandford--Znajek solution.  This is the curved-space analogue
of the Bogovalov construction \cite{Bogovalov:1999vg}. Since it holds that $d \theta  \wedge  d \psi  = \mathcal{O}  ( a_\star^2  ) $, one can simply characterize the inclined current sheet in terms of $\theta $ at zeroth order in spin. Accounting for corrections at $ \mathcal{O}  ( a_\star^2  ) $, it is convenient to define
\begin{equation}
 \cos\vartheta
 =
 1-\frac{\psi}{\psi_0}  , 
 \label{eq:appB_magnetic_colatitude}
\end{equation}
for which it holds that $d \vartheta  \wedge d \psi =0  $. A split whose normal is inclined by an angle $\chi$ with respect to
the black-hole spin axis is then described in field-sheet coordinates
by the tilted surface
\begin{equation}
 \Xi_{\rm BZ}
 =
 \cos\chi\cos\vartheta
 +
 \sin\chi\sin\vartheta\cos\Phi .
 \label{eq:appB_exact_split}
\end{equation}
Because $\vartheta$ is a function of $\psi$, Eq.~\eqref{eq:appB_exact_split}
depends only on the two Euler labels $(\psi,\Phi)$ and therefore obeys
Eq.~\eqref{eq:appB_no_monopole} identically.  The radial dependence of
$\Phi$ produces the retarded, striped structure of the oblique sheet,
whereas the finite-spin deformation of the Blandford-Znajek flux surfaces produces
an additional radial warp (see Fig.~\ref{fig:warped_sheet}). Using Eq.~\eqref{eq:BZ_flux}, we find
\begin{equation}
 \cos\vartheta
 =
 \cos\theta
 -
 a_\star^2\Rhat(r/M)\sin^2\theta\cos\theta
 +O(a_\star^4),
\end{equation}
and hence
\begin{equation}
 \vartheta
 =
 \theta
 +
 a_\star^2\Rhat(r/M)\sin\theta\cos\theta
 +O(a_\star^4).
 \label{eq:appB_vartheta_expansion}
\end{equation}
Let $\theta_{\rm B}(\Phi)$ denote the corresponding Bogovalov
sheet,
\begin{equation}
 \cos\chi\cos\theta_{\rm B}
 +
 \sin\chi\sin\theta_{\rm B}\cos\Phi
 =0  . 
 \label{eq:appB_Michel_sheet}
\end{equation}
Because $ \vartheta_{\Sigma} $ and $\theta_{\rm B}$ obey an identical constraint equation, it holds that
\begin{equation}
 \vartheta(r,\theta_\Sigma)
 =
 \theta_{\rm B}(\Phi).
\end{equation}
Substitution of Eq.~\eqref{eq:appB_vartheta_expansion} gives
\begin{equation}
 \theta_\Sigma(r,\Phi)
 =
 \theta_{\rm B}(\Phi)
 -
 a_\star^2\Rhat(r/M)
 \sin\theta_{\rm B}(\Phi)
 \cos\theta_{\rm B}(\Phi)
 +O(a_\star^4).
 \label{eq:appB_warped_sheet}
\end{equation}
For the alignment calculation we require only the small-inclination limit.  For $\chi\ll1$,
\begin{equation}
 \theta_{\rm B}
 =
 \frac{\pi}{2}+\chi\cos\Phi+O(\chi^2),
\end{equation}
so Eq.~\eqref{eq:appB_warped_sheet} reduces to
\begin{equation}
 \theta_\Sigma
 =
 \frac{\pi}{2}
 +
 \chi\left[1+a_\star^2\Rhat(r/M)\right]\cos\Phi
 +O(\chi^2,a_\star^4).
 \label{eq:appB_small_chi_sheet}
\end{equation}
The same reversal surface can be described geometrically, to this order, by
\begin{equation}
 S
 =
 \theta-\frac{\pi}{2}
 -
 \chi\left[1+a_\star^2\Rhat(r/M)\right]\cos\Phi .
 \label{eq:appB_sheet_argument}
\end{equation}
Only the zero of $S$ is used here. The smooth reversal is $\mathcal Q(\Xi_{\rm BZ})$, so all derivatives entering the diffusion calculation act on the full field-line label before evaluation on the sheet.

\subsection{Compatibility with the magnetic constraint}
We verify the magnetic Gauss law throughout the smooth layer, rather than only on the reversal surface. For
\begin{equation}
 B^i=\mathcal Q(\Xi_{\rm BZ})B^i_{\rm ext},\qquad
 \partial_i(\sqrt\gamma B^i_{\rm ext})=0,
\end{equation}
the constraint becomes
\begin{equation}
 \partial_i(\sqrt\gamma B^i)
 =\sqrt\gamma\,\mathcal Q'(\Xi_{\rm BZ})B^i_{\rm ext}\partial_i\Xi_{\rm BZ}.
 \label{eq:divB_common_reversal_check}
\end{equation}
On a spatial slice, the Euler-potential representation gives
\begin{equation}
 \sqrt\gamma B^i_{\rm ext}=[ijk]\,\partial_j\psi\,\partial_k\Phi,
 \qquad [r\theta\phi]=1,
\end{equation}
where $[ijk]$ is the alternating symbol. Hence
\begin{equation}
 B^i_{\rm ext}\partial_i\psi=0,\qquad
 B^i_{\rm ext}\partial_i\Phi=0.
 \label{eq:BdotgradS_before_sheet}
\end{equation}
Since $\vartheta=\vartheta(\psi)$ and $\Xi_{\rm BZ}=\Xi_{\rm BZ}(\psi,\Phi;\chi(t))$, its spatial derivative obeys
\begin{equation}
 B^i_{\rm ext}\partial_i\Xi_{\rm BZ}
 =\frac{\partial\Xi_{\rm BZ}}{\partial\psi}B^i_{\rm ext}\partial_i\psi
 +\frac{\partial\Xi_{\rm BZ}}{\partial\Phi}B^i_{\rm ext}\partial_i\Phi=0.
 \label{eq:BdotgradS_zero}
\end{equation}
Consequently,
\begin{equation}
 \partial_i(\sqrt\gamma B^i)=0
\label{eq:divB_common_reversal_result} ~~ . 
\end{equation}

\section{Flux decay}
\label{eq:fluxdecay}
\begin{figure}
 \centering
\includegraphics[width=0.6\linewidth] {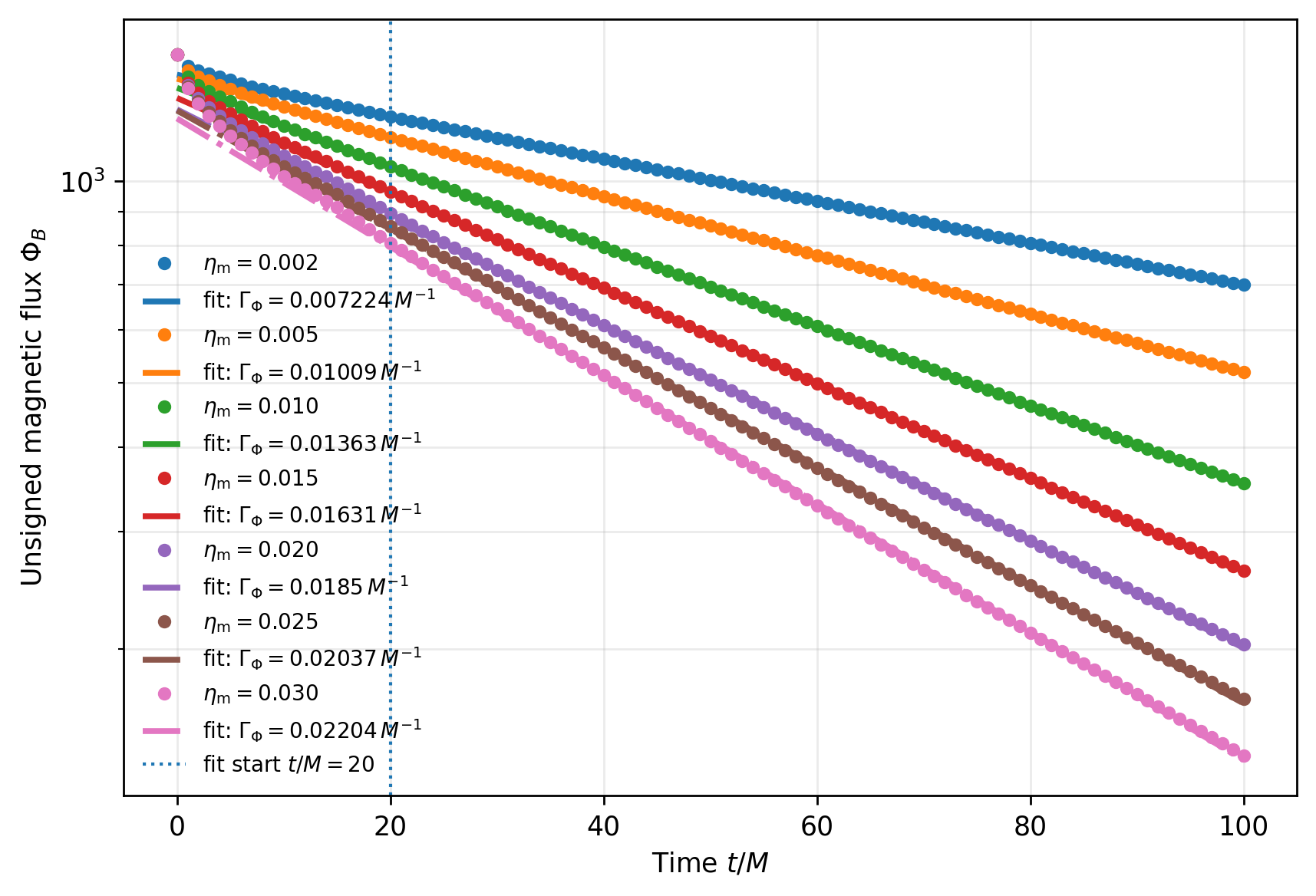}
 \caption{\label{fig:flux_decay}
 Late-time decay of the magnetic flux $\Phi_B$ for the
 resistivity scan.  Symbols show the numerical data and dash-dotted lines
 exponential fits of Eq.~\eqref{eq:main_flux_decay}, performed for
 $t/M\geq20$ (vertical dotted line).}
\end{figure}

\begin{figure}
 \centering
 \includegraphics[width=0.6\linewidth]
 {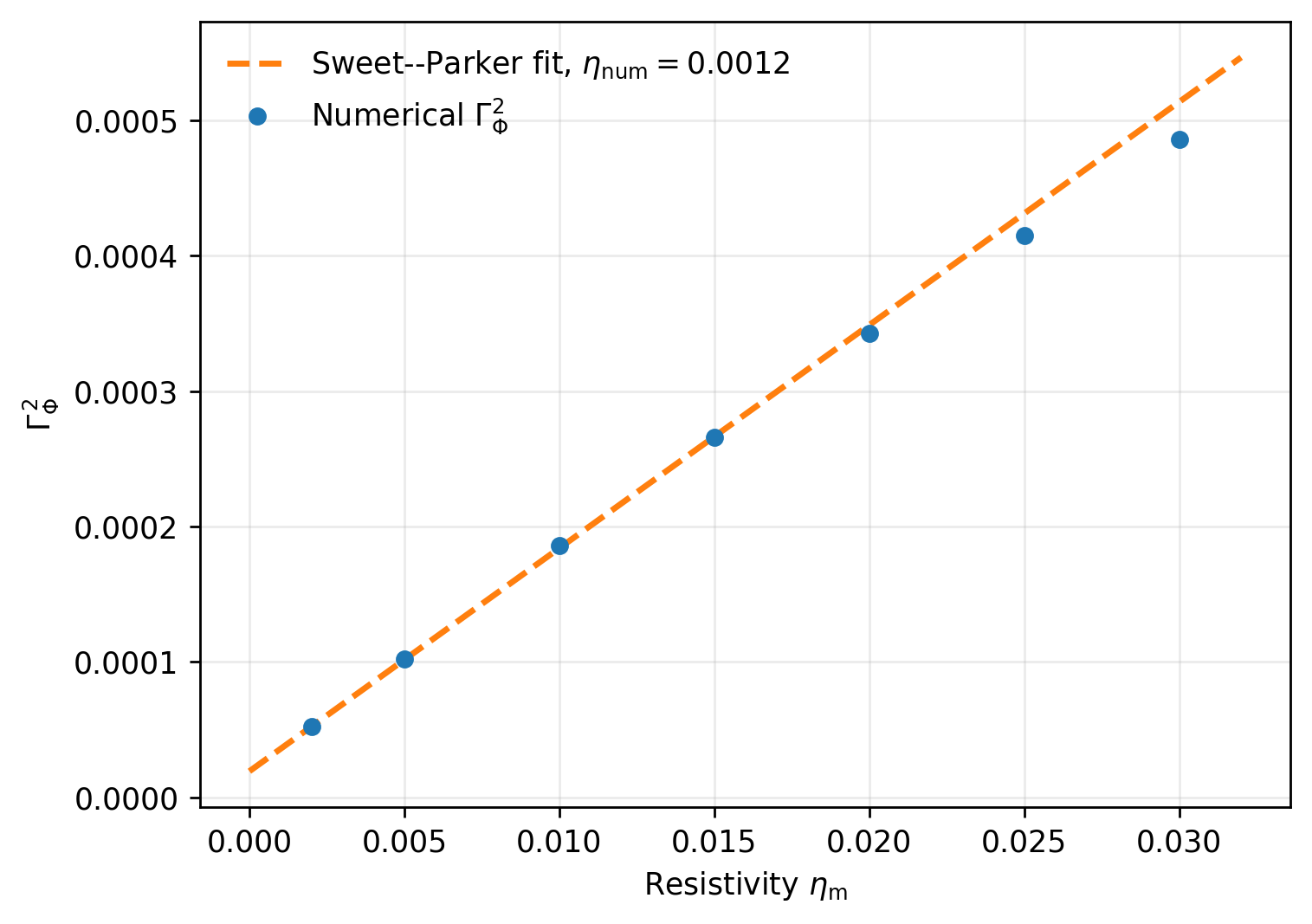}
 \caption{\label{fig:flux_eta}
 Squared magnetic-flux decay rate as a function of the bare
 resistivity.  Sweet--Parker reconnection predicts
 $\Gamma_\Phi\propto\eta_{\rm m}^{1/2}$, and hence
 $\Gamma_\Phi^2$ linear in
 $\eta_{\rm m}$.
 The dashed line is a linear fit to the numerical rates and gives
 $\eta_{\rm num}^{(\Phi)}/M=0.0012$.}
\end{figure}

In this section we infer numerical resistivity from the flux decay by finding the offset needed to obtain a square-root resistivity dependence of flux decay. We monitor the magnetic flux
$\Phi_B$ threading the black hole given by 
\begin{align}
    \Phi_B(t)
    =
    \frac{1}{2}
    \int_{r=r_{\rm H}}
    \left|B^r\right|
    \sqrt{\gamma}\,
    \mathrm d\theta\,\mathrm d\phi .
    \label{eq:main_unsigned_flux}
\end{align}
Once the initial transient has
passed, it is accurately described by
\begin{equation}
 \Phi_B(t)\propto\exp(-\Gamma_\Phi t).
 \label{eq:main_flux_decay}
\end{equation}
Figure~\ref{fig:flux_decay} shows exponential fits beginning at
$t/M=20$. 
In these simulations the reconnecting layer remains in the regime of Sweet--Parker reconnection \cite{Sweet_1958,Parker1957}.  Its reconnection rate therefore scales as
$\eta_{\rm effective }^{1/2}$, with $ \eta_{\rm effective }  = \eta_{\rm m}+\eta_{\rm num}^{(\Phi)} $. This implies
\begin{equation}
 \Gamma_\Phi^2
 =
 C\left(
 \eta_{\rm m}+\eta_{\rm num}^{(\Phi)}
 \right)
 \label{eq:main_SP_fit}
\end{equation}
for a constant $C$.  Figure~\ref{fig:flux_eta} shows that the measured rates satisfy this linear relation accurately at low resistivity.  The
intercept gives
\begin{equation}
 \eta_{\rm num}^{(\Phi)}
 \simeq 1.2\times10^{-3}M .
 \label{eq:main_flux_eta_num}
\end{equation}

\section{Convergence test}
\label{eq:convergence}
\begin{figure}
    \centering
    \includegraphics[width=0.5\linewidth]{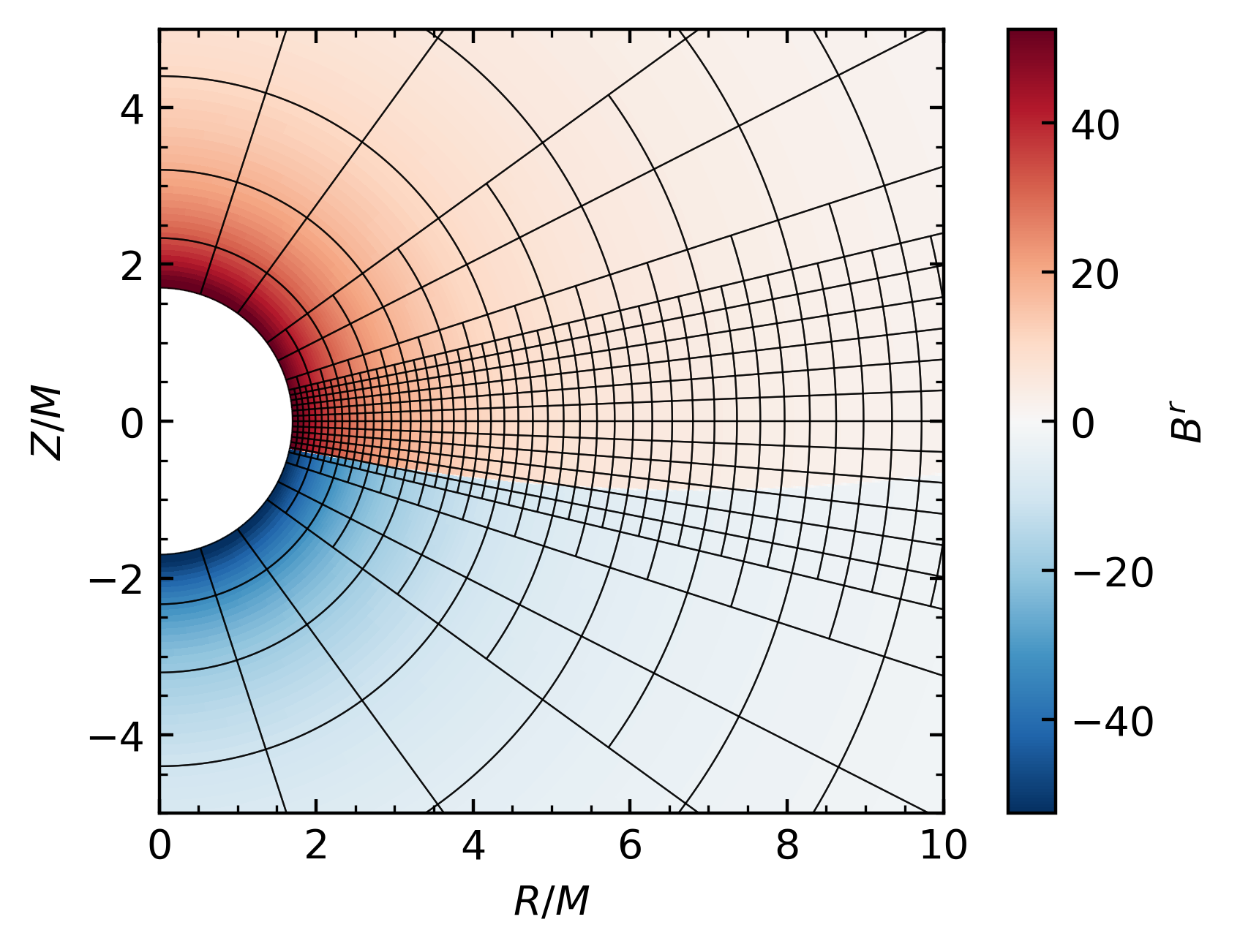}
    \caption{\label{fig:conv_amr}
    Initial $B^r$ on the $\phi=0$ meridional slice for the
    high-resolution convergence run, with the static mesh-refinement
    blocks overlaid. Relative to the standard grid, the base
    grid is unchanged but one additional refinement layer is added
    around the tilted current sheet, so that the polarity reversal is
    enclosed by a finer highest-refinement region over the displayed
    domain.}
\end{figure}

To assess the sensitivity of the measured alignment rate to numerical
resolution, we repeat a representative run at fixed spin and
resistivity, namely $a_*=0.6$ and $\eta_{\rm m}/M=0.01$, using a
higher resolution static-mesh-refinement setup (see Fig.~\ref{fig:conv_amr}). In this convergence run, the base
grid is kept unchanged, while an additional refinement layer is added
around the current sheet.

\begin{figure}
    \centering
    \includegraphics[width=0.5\linewidth]{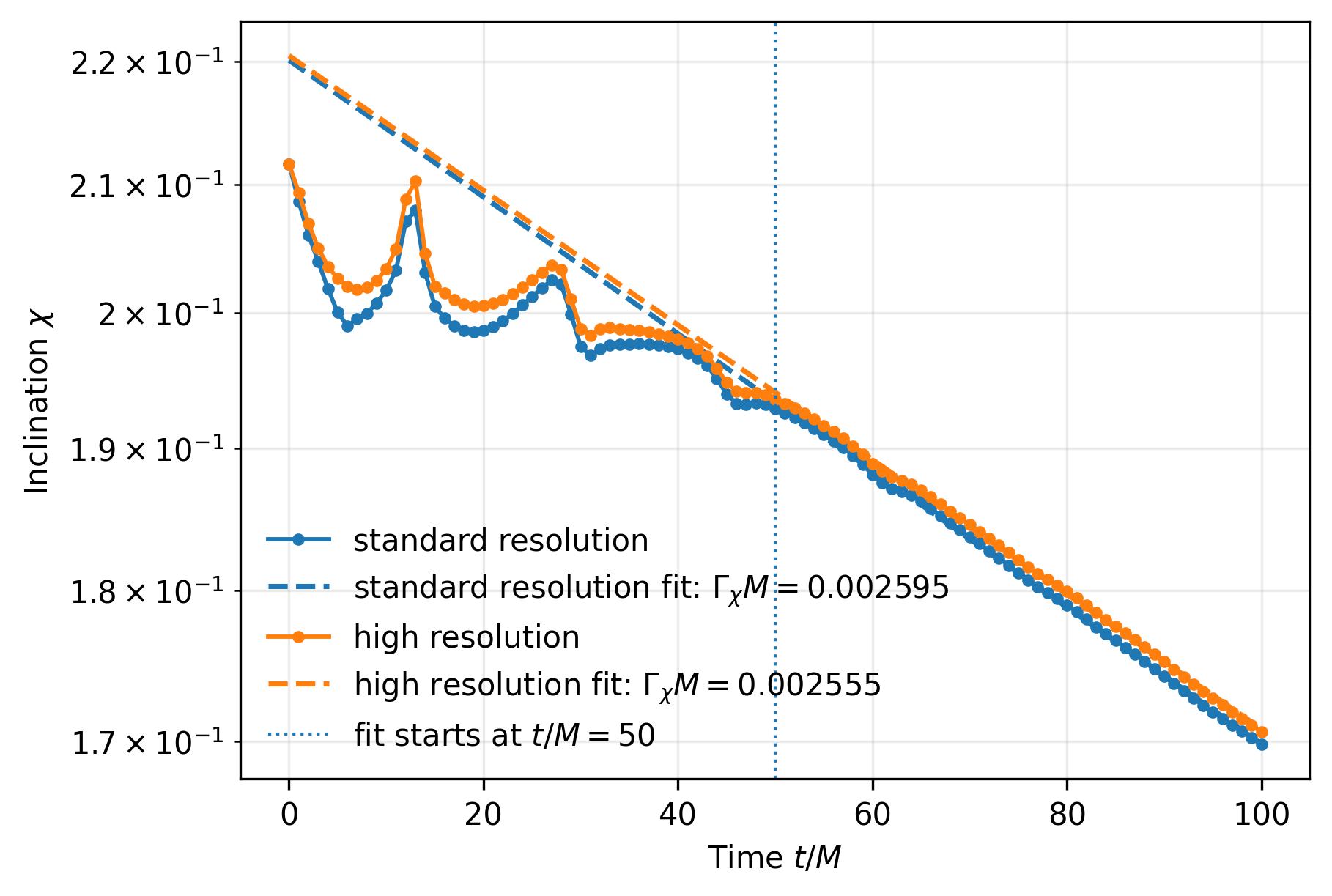}
    \caption{\label{fig:conv_alignment}
    Convergence test for the alignment history at $a_*=0.6$ and
    $\eta_{\rm m}/M=0.01$. The standard-resolution (effective $400^3$ cells) and a double-resolution (effective $800^3$ cells)
    run is shown together with late-time exponential fits. The fitted
    decay rates are very similar, showing that the secular alignment
    rate changes only weakly when the refinement around the current
    sheet is increased.}
\end{figure}

Figure~\ref{fig:conv_alignment} compares the inclination
$\chi(t)$ for the standard- and high-resolution runs, together with
late-time exponential fits of the form
$\chi(t)=A e^{-\Gamma_\chi t}$. The two curves show very similar
late-time decay, and the fitted alignment rates are nearly the same.
This indicates that, for this run, the secular alignment rate is only
weakly affected by the additional refinement. In particular, the
resolution used in the main text already captures the dominant
late-time alignment behavior well.

 \end{document}